\documentclass[twocolumn]{aastex701}

\newcommand{\psrseven}{PSR~J0740$+$6620} 
\newcommand{\nicer}{\textit{NICER}}
\newcommand{\Reqs}{R_{\rm eq}}
\defcitealias{Kalapotharakos2021}{CK21}
\defcitealias{Olmschenk2025}{GO25}

\begin{document}

\title{Multipolar Magnetic-Field Inference for PSR J0740+6620 \\
with Neural-Network-Accelerated NICER Pulse-Profile Modeling}

\author[orcid=0000-0003-3481-5913,sname=Taiyebah]{Farhana Taiyebah}
\affiliation{Theoretical Division, Los Alamos National Laboratory, Los Alamos, NM 87545, USA}
\affiliation{Department of Scientific Computing, Florida State University, Tallahassee, FL 32306, USA}
\email{ftaiyebah@fsu.edu}

\author[orcid=0000-0003-1080-5286, sname=Kalapotharakos]{Constantinos Kalapotharakos} 
\affiliation{Astrophysics Science Division, NASA Goddard Space Flight Center, Greenbelt, MD 20771, USA}
\email{konstantinos.kalapotharakos@nasa.gov}

\author[orcid=0000-0001-8472-2219, sname=Olmschenk]{Greg Olmschenk}
\affiliation{Astrophysics Science Division, NASA Goddard Space Flight Center, Greenbelt, MD 20771, USA}
\email{greg@olmschenk.com}

\author[orcid=0009-0002-9936-4041, sname=Wallace]{Wendy F. Wallace}
\affiliation{Hamburger Sternwarte, Universität Hamburg, Gojenbergsweg 112, D-21029 Hamburg, Germany}
\affiliation{Astrophysics Science Division, NASA Goddard Space Flight Center, Greenbelt, MD 20771, USA}
\email{wendy.wallace@uni-hamburg.de}

\author[orcid=0000-0002-3316-5149, sname=De]{Soumi De}
\affiliation{Theoretical Division, Los Alamos National Laboratory, Los Alamos, NM 87545, USA}
\affiliation{Center for Theoretical Astrophysics, Los Alamos National Laboratory, Los Alamos, NM 87545, USA}
\email{soumide@lanl.gov}

\author[orcid=0009-0009-4871-7231 ,sname=Siddik]{Abu Bucker Siddik}
\affiliation{Computing and Artificial Intelligence Division, Los Alamos National Laboratory, Los Alamos,  NM 87545, USA}
\email{siddik@lanl.gov}

\author[orcid=0000-0002-1353-3688 ,sname=Oyen]{Diane Oyen}
\affiliation{Computing and Artificial Intelligence Division, Los Alamos National Laboratory, Los Alamos,  NM 87545, USA}
\email{doyen@lanl.gov}

\author[orcid=0000-0001-6284-2842, sname=Lechien]{Thibault Lechien}
\affiliation{Max Planck Institute for Astrophysics, Karl-Schwarzschild-Straße 1, 85748 Garching bei München, Germany}
\affiliation{Astrophysics Science Division, NASA Goddard Space Flight Center, Greenbelt, MD 20771, USA}
\email{thibaultlechien@gmail.com}

\author[orcid=0000-0002-9249-0515, sname=Wadiasingh]{Zorawar Wadiasingh}
\affiliation{Department of Astronomy, University of Maryland, College Park, MD 20742, USA}
\affiliation{Astrophysics Science Division, NASA Goddard Space Flight Center, Greenbelt, MD 20771, USA}
\affiliation{Center for Research and Exploration in Space Science and Technology, NASA/GSFC, Greenbelt, MD 20771, USA}
\email{zwadiasingh@gmail.com}



\begin{abstract}
We investigate the multipolar surface magnetic-field structure of the high-mass millisecond pulsar PSR~J0740$+$6620 using the 32-bin bolometric \nicer\ pulse profile of \citet{Dittmann2024}. Building on the neural-network surrogate framework of \citet{Olmschenk2025}, we model the emitting regions as open-field-line footpoints of an offset dipole plus axisymmetric quadrupole static-vacuum field, rather than as prescribed geometric hotspots. We fix the stellar mass, radius, observer inclination, and hotspot temperature ratio to the \citet{Dittmann2024} maximum-likelihood values and explore the resulting 11-dimensional magnetic-field space. To make this feasible, we train convolutional neural-network surrogates on $5.12\times10^7$ synthetic bolometric light curves and use them in a parallel ensemble Markov Chain Monte Carlo (MCMC) calculation on 4000 CPU cores, accelerating likelihood evaluations by a factor of $\gtrsim 400$.
We perform independent inferences for two calibrated temperature-weight prescriptions, $T_w=1.31$ and $T_w=1.41$, encoding the relative bolometric weight associated with the hotspot temperature difference. The posteriors, posterior-predictive light curves, and maximum-likelihood footpoints are very similar, indicating weak sensitivity to this choice. The offset model reproduces the observed double-peaked profile and yields broad, multimodal posteriors, reflecting both the background-dominated data and degeneracies of the multipolar parameterization. The hotspot-density map shows that pulse phases constrain the approximate azimuthal placement of the emission, while latitude, surface extent, and morphology remain weakly constrained. A restricted zero-offset run is disfavored within the adopted field basis. This work extends neural-network-accelerated magnetic-field inference to PSR~J0740$+$6620 and motivates future energy-dependent, force-free, and joint X-ray/$\gamma$-ray extensions.
\end{abstract}



\section{Introduction}\label{sec:intro}

Rotation-powered millisecond pulsars (MSPs) are neutron stars that have been spun up to millisecond rotation periods through long-term accretion from a binary companion \citep{Alpar1982,BhattacharyaVDH1991}. Their thermal soft-X-ray emission is generally interpreted as originating from magnetic polar caps heated by returning magnetospheric particles, either through pair-cascade activity above the polar caps \citep{HardingMuslimov2001,HardingMuslimov2002} or through return currents associated with the global pulsar magnetosphere \citep{Contopoulos1999,BaiSpitkovsky2010, Kalapotharakos2014}. Because the observed X-ray pulse profiles depend sensitively on the geometry of the emitting regions and on general-relativistic photon propagation near the stellar surface, detailed pulse-profile modeling provides a powerful probe of neutron-star structure and magnetic-field topology.

Observations by the {\it Neutron Star Interior Composition Explorer} ({\it NICER}; \citealt{Gendreau2016}) have transformed this field by enabling high-precision X-ray waveform measurements for nearby MSPs. The first major {\it NICER} target, PSR~J0030$+$0451 \citep{Bogdanov2019}, revealed hotspot configurations inconsistent with a centered dipolar magnetic field. Independent analyses by \citet{Miller2019} and \citet{Riley2019} inferred that both dominant emitting regions lie within the same rotational hemisphere, implying the presence of substantial non-dipolar magnetic structure \citep{Bilous2019}. This result motivated a series of studies investigating increasingly complex magnetic-field configurations for MSPs \citep{Chen2020,Kalapotharakos2021,Petri2023}.

\citet[][hereafter \citetalias{Kalapotharakos2021}]{Kalapotharakos2021} developed a physical pulse-profile modeling framework in which the near-surface magnetic field is represented as the superposition of an offset dipole and an offset $m=0$ quadrupole. In this approach, the footpoints of the open magnetic field lines define
the hotspot geometry, while the observed X-ray pulse profile is computed 
with the \textsc{GIKS} code using relativistic ray tracing in a Schwarzschild spacetime, including rotational Doppler boosting and the corresponding relativistic transformation of the emitted intensity, as detailed in \citetalias{Kalapotharakos2021}.\footnote{The \textsc{GIKS} code also supports ray tracing in the Kerr metric. However, Kerr does not describe the exact exterior spacetime of a rotating neutron star. For the present application, frame-dragging effects are expected to be subdominant, and we therefore adopt the Schwarzschild-plus-Doppler approximation.} The corresponding parameter space contains eleven free magnetic-field parameters, and the resulting likelihood surface exhibits substantial degeneracy. Although the \citetalias{Kalapotharakos2021} framework demonstrated that multipolar magnetic fields can reproduce the observed {\it NICER} pulse profile of PSR~J0030$+$0451, the computational cost of the physical forward model severely limited the achievable depth of MCMC exploration.

\citet[][hereafter \citetalias{Olmschenk2025}]{Olmschenk2025} addressed this computational bottleneck by introducing a convolutional neural network (NN) surrogate model trained directly on synthetic light curves generated by the \citetalias{Kalapotharakos2021} physical model. Their neural-network surrogate reproduced the physical model with high fidelity while accelerating likelihood evaluations by more than a factor of $\sim$400. This acceleration enabled substantially more complete MCMC exploration of the PSR~J0030$+$0451 parameter
space. Importantly, \citetalias{Olmschenk2025} demonstrated that the earlier MCMC runs of \citetalias{Kalapotharakos2021} had not fully reached equilibrium, and that the resulting posterior distributions differed in physically meaningful ways from the previously inferred solutions.

The high-mass MSP PSR~J0740$+$6620 represents both a natural extension and a substantially more demanding test of this methodology. With a precisely measured gravitational mass of $2.08\pm0.07\,M_\odot$ \citep{Cromartie2020,Fonseca2021} and NICER+XMM-Newton radius measurements \citep{Miller2021,Riley2021,Dittmann2024,Salmi2024}, PSR~J0740$+$6620 is an important observational anchor for dense-matter equation-of-state studies \citep{Miller2021,Raaijmakers2021}. Its X-ray pulse profile has been modeled independently by \citet{Miller2021} and \citet{Riley2021}, and more recently by
\citet{Dittmann2024} and \citet{Salmi2024} using 3.6\,yr of {\it NICER} data combined with {\it XMM-Newton} observations. These analyses inferred hotspot configurations that depart from the antipodal geometry expected for a centered dipolar magnetic field.

However, PSR~J0740$+$6620 is observationally much more challenging than PSR~J0030$+$0451. The {\it NICER} count rate is only $\sim$5\% of that of PSR~J0030$+$0451 \citep{Miller2021}, and the pulsed signal is superposed on a substantially larger background contribution. After background
subtraction, the pulsed counts span only $\sim$666--2545 counts per phase bin, while the background
contributes $\sim$30,398 counts per bin. Consequently, the per-bin photon-noise uncertainty is comparable to a large fraction of the pulsed amplitude. This lower signal-to-noise ratio broadens the volume of parameter space consistent with the data, thereby increasing the range of magnetic-field configurations allowed by the intrinsic model degeneracies and broadening the posterior distributions, making long MCMC chains essential for robust posterior inference.

In this work, we carry out a first exploration of the multipolar magnetic-field structure of PSR~J0740$+$6620 using static vacuum field configurations. We adopt from \citet{Dittmann2024} the maximum-likelihood values of the stellar mass, radius, observer inclination, and the temperature ratio between the two emitting regions, and hold these quantities fixed throughout the present analysis. Under these assumptions, we explore the multipolar magnetic-field parameter space and determine which field structures can reproduce the observed X-ray pulse profile.

Although static vacuum fields do not provide a fully self-consistent description of the plasma-filled pulsar magnetosphere, they offer a useful first approximation for connecting surface magnetic structure to polar-cap and hotspot morphology. Force-free magnetospheres include self-consistent charge and current distributions and generally produce polar caps that are larger and modified relative to the static-vacuum case. Nonetheless, comparisons with force-free calculations suggest that the locations and broad morphology of the polar caps remain closely related to those obtained from static vacuum fields \citep{Kalapotharakos2021}. This connection will be examined further by \citet{Lechien2026}. We therefore use the static-vacuum framework as a controlled first step toward more realistic force-free magnetic-topology inference.

To make this exploration computationally feasible, we generate a training database of $\sim$$5.12\times10^{7}$ synthetic light curves using the physical forward model, train a modified 32-phase-bin convolutional neural-network surrogate adapted from the \citetalias{Olmschenk2025} architecture, and integrate the trained surrogate into a large-scale parallel MCMC framework. This neural-network-accelerated approach allows us to explore the highly degenerate multipolar parameter space much more extensively than would be possible with direct physical-model evaluations alone. Without the surrogate, an MCMC run of equivalent depth would require approximately 2 years of continuous computation on the same 4000-core system --- far beyond what is practical --- meaning the neural-network surrogate is not merely an optimization but a prerequisite for this analysis. This work constitutes the first application of the \citetalias{Olmschenk2025} surrogate-based methodology to a {\it NICER} millisecond pulsar other than PSR~J0030$+$0451 and provides a practical step toward future joint X-ray/$\gamma$-ray magnetic-topology inference with more realistic force-free magnetospheres.


The paper is organized as follows. Section~\ref{sec:target} summarizes the observational data and adopted stellar parameters for PSR~J0740$+$6620. Section~\ref{sec:model_setup} describes the adopted source parameters and the calibration of the simplified bolometric emission prescription against the \citet{Dittmann2024} maximum-likelihood reference light curve. Section~\ref{sec:phys_model} reviews the physical forward model and the adopted temperature-asymmetry prescription. Section~\ref{sec:nn} presents the generation of the training
database, the network architecture, and the validation tests. Section~\ref{sec:mcmc} describes the neural-network-accelerated MCMC implementation and convergence diagnostics. Section~\ref{sec:results} presents the posterior distributions, predicted posterior 
light curves, and hotspot density maps. Section~\ref{sec:discussion} discusses the implications of the results and future extensions toward joint X-ray/$\gamma$-ray and force-free inference. Section~\ref{sec:conclusions} summarizes our conclusions.

\section{Target and Observational Data}\label{sec:target}

\subsection{PSR~J0740$+$6620}\label{sec:target_params}

PSR~J0740$+$6620 is a $346.53$\,Hz millisecond pulsar in a binary system with a white-dwarf companion. Radio-timing measurements of the Shapiro delay yield a high-precision gravitational mass of $M = 2.08\pm0.07\,M_\odot$ \citep{Cromartie2020,Fonseca2021}, making PSR~J0740$+$6620 one of the most massive neutron stars currently known. The source has been studied with {\it NICER} pulse-profile modeling by \citet{Miller2021} and \citet{Riley2021}, and more recently by \citet{Dittmann2024} using a combined {\it NICER}+{\it XMM-Newton} analysis based on 3.6\,yr of observations.

The spin period is $P = 1/346.53\ {\rm s}$, corresponding to a light-cylinder radius
\begin{equation}
R_{\rm LC} = \frac{cP}{2\pi} \approx 138\ {\rm km}.
\label{eq:rlc}
\end{equation}
For the stellar radius adopted below, this corresponds to $R_{\rm LC}\approx 12 \Reqs$. The relatively small ratio of $R_{\rm LC}/\Reqs$ is relevant for the open-field-line calculation described in Section~\ref{sec:phys_model}.

\subsection{{\it NICER} bolometric pulse profile}\label{sec:target_data}

The bolometric {\it NICER} pulse profile used in this work is taken from \citet{Dittmann2024} and consists of $N_{\rm bin}=32$ phase bins. This profile incorporates the source and background treatment from their joint {\it NICER}+{\it XMM-Newton} analysis. The total observed counts per phase bin, denoted $T_i$, lie in the range $\sim$$3.11\times 10^{4}$--$3.29\times 10^{4}$, with a mean background contribution of $B=30,398$ counts per phase bin. After subtraction of the background, the pulsed signal, $D_i=T_i-B$, spans only $\sim$666--2545 counts per bin across rotational phase.

Assuming Poisson uncertainties for both the total counts and the background contribution, the uncertainty in each background-subtracted phase bin is
\begin{equation}
\sigma_i = \sqrt{T_i+B}
          = \sqrt{D_i+2B},
\label{eq:sigma}
\end{equation}
following the prescription adopted in \citetalias{Kalapotharakos2021}. The resulting uncertainties are approximately $\sigma_i\approx 248$--$252$ counts per phase bin.

\begin{figure}[ht]
  \centering
  \includegraphics[width=\columnwidth]
  {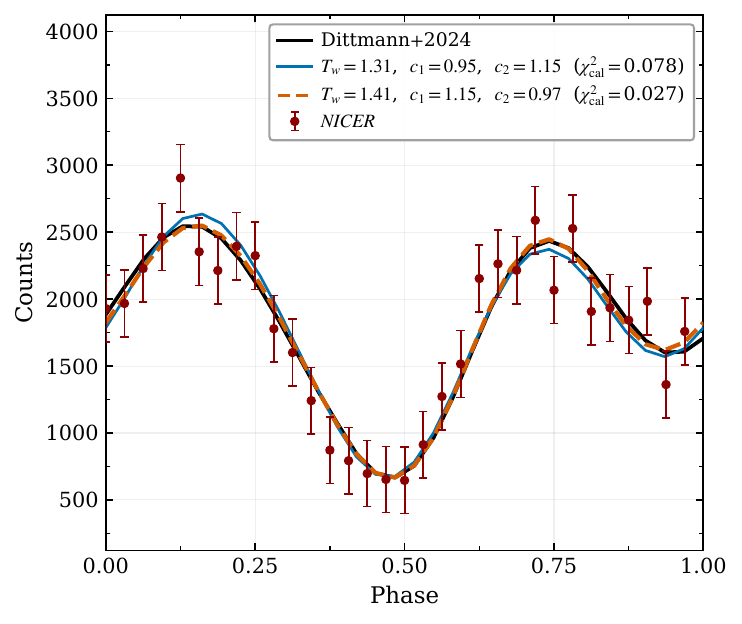}
  \caption{%
  Calibration of the simplified bolometric light-curve model. Red points show the background-subtracted \nicer\ pulse profile of \psrseven\ with $1\sigma$ uncertainties. The black curve is the \citet{Dittmann2024} maximum-likelihood bolometric reference model. The blue and orange curves show our calibrated forward-model light curves for $T_w=1.31$ and $T_w=1.41$, respectively. Both calibrations reproduce the reference profile closely and are used as fixed emission prescriptions in the magnetic-field inference.}
  \label{fig:lightcurve_beaming}
\end{figure}

Thus, the bolometric profile is measured in a background-dominated regime: although the pulsed signal is clearly detected, the statistical uncertainty in each phase bin is a significant fraction of the background-subtracted pulsed counts. Figure~\ref{fig:lightcurve_beaming} shows the background-subtracted bolometric pulse profile with the corresponding uncertainties used throughout this work.

\section{Model Setup and Emission Calibration}\label{sec:model_setup}

\subsection{Adopted source parameters and bolometric emission calibration}
\label{sec:emission_calibration}

In the present analysis, we do not refit the stellar or observer-geometry parameters. Instead, we adopt from \citet{Dittmann2024} the maximum-likelihood values of the parameters that define the relativistic ray-tracing setup. Specifically, we fix the gravitational mass to $M = 1.958\,M_\odot$, the equatorial radius to $\Reqs = 11.541\,{\rm km}$, and the observer inclination to $\theta_{\rm obs}=1.539\,{\rm rad}$. These quantities determine the photon library used in the light-curve calculations below.

We also adopt the maximum-likelihood effective temperatures of the two emitting regions reported by \citet{Dittmann2024}, $kT_{{\rm eff},1}=0.106\,{\rm keV}$ and $kT_{{\rm eff},2}=0.099\,{\rm keV}$. Since the present work models the bolometric pulse profile, the absolute temperatures enter our treatment primarily through the relative bolometric luminosity weight of the two emitting regions. The nominal value suggested by the temperature ratio is therefore $T_w=(T_{{\rm eff},1}/T_{{\rm eff},2})^4\simeq1.31$, with the hotter region assigned the larger weight.

Following the spirit of \citetalias{Kalapotharakos2021}, we use a simplified angular-emission prescription rather than reproducing the full atmosphere model used in the \citet{Dittmann2024} pulse-profile analysis. In the ray-tracing calculation, each photon trajectory that intersects an emitting region is assigned an additional luminosity weight
\begin{equation}
\mathcal{W}_k =
\left(\frac{\eta_k}{1+z_s}\right)^3
\left(\eta_k\cos\alpha_k\right)^{c_k}
W_k ,
\label{eq:photon_weight}
\end{equation}
where $\eta_k = \sqrt{1-u^2}/(1-u\cos\xi_k)$ is the Doppler factor, with $u$ the co-rotation speed of the emitting surface element (in units of $c$) and $\xi_k$ the angle between the element's velocity and the photon propagation direction; $z_s$ is the surface gravitational redshift, $\alpha_k$ is the photon emission angle relative to the outward surface normal in the comoving frame, and $c_k$ is the limb darkening exponent of the simplified atmosphere model. The factor $W_k$ encodes the relative bolometric luminosity of the two emitting regions: $W_k=T_w$ for photons originating from the higher-temperature hotspot and $W_k=1$ for photons originating from the lower-temperature hotspot. Because the two emitting regions in the \citet{Dittmann2024} maximum-likelihood solution have different temperatures, we allow different beaming exponents, denoted $c_1$ and $c_2$, for the two regions.

The overall normalization of the model light curve is not fixed directly by the physical distance to the source. Instead, as in \citetalias{Kalapotharakos2021}, the photon weights are normalized empirically by requiring the simplified model to reproduce a reference bolometric light curve. In the present case, the reference is the maximum-likelihood bolometric model light curve of \citet{Dittmann2024}. For this calibration step only, we use their maximum-likelihood circular-hotspot geometry. The primary hotspot is centered at colatitude $\theta_{c1}=1.387\,{\rm rad}$ with angular radius $\rho_1=0.092\,{\rm rad}$, while the secondary hotspot is centered at colatitude $\theta_{c2}=1.980\,{\rm rad}$ with angular radius $\rho_2=0.112\,{\rm rad}$. The relative phase offset between the two hotspots is $\delta_{\rm phase}=0.428$ cycles. These circular hotspots are used only to calibrate the simplified emission prescription and are not imposed in the magnetic-field inference.

We perform two related calibrations. First, we fix the temperature-weight factor to its nominal bolometric value, $T_w=1.31$, and determine the values of $c_1$ and $c_2$, together with the overall normalization and phase alignment, that best reproduce the \citet{Dittmann2024} maximum-likelihood bolometric reference light curve. Second, we allow $T_w$, $c_1$, and $c_2$ to vary together in the calibration. This gives a best-fitting effective temperature-weight factor of $T_w\simeq1.41$, slightly larger than the nominal value derived from the temperature ratio through the $T^4$ bolometric scaling.

For each calibration trial, the agreement with the reference light curve is quantified using
\begin{equation}
\chi^2_{\rm cal} =
\sum_{i=1}^{N_{\rm bin}}
\frac{\left(L^{\rm ref}_i-L^{\rm cal}_i\right)^2}{\sigma_i^2},
\label{eq:chi2_cal}
\end{equation}
where $L^{\rm ref}_i$ is the \citet{Dittmann2024} maximum-likelihood bolometric reference light curve, $L^{\rm cal}_i$ is the corresponding light curve generated by our simplified prescription using the circular reference hotspots, and $\sigma_i$ is the observational uncertainty defined in Equation~\ref{eq:sigma}. We use these observational uncertainties because the purpose of the calibration is to match the reference model on the same count scale and with the same phase-bin weighting as the observed bolometric profile. The quantity $\chi^2_{\rm cal}$ is used only to calibrate the simplified emission prescription and should not be confused with the likelihood used later for the magnetic-field inference.

\begin{table}[ht]
\centering
\caption{Calibration of the simplified bolometric emission prescription against the \citet{Dittmann2024} maximum-likelihood reference light curve. The case $T_w=1.31$ fixes the relative luminosity weight to the value implied by the reference-model temperature ratio, while the case $T_w=1.41$ corresponds to the best effective luminosity weight obtained when $T_w$, $c_1$, and $c_2$ are varied together. The final column gives the calibration mismatch defined in Equation~\ref{eq:chi2_cal}.}
\label{tab:beaming_calibration}
\begin{tabular}{cccc}
\hline\hline
$T_w$ & $c_1$ & $c_2$ & $\chi^2_{\rm cal}$ \\
\hline
$1.31$ & $0.95$ & $1.15$ & 0.078 \\
$1.41$ & $1.15$ & $0.97$ & 0.027 \\
\hline
\end{tabular}
\end{table}

After this calibration, the values of $T_w$, $c_1$, $c_2$, the normalization, and the phase alignment are held fixed for the corresponding magnetic-field inference run. Thus, the calibration parameters define the simplified bolometric emission model, while the MCMC itself explores only the magnetic-field parameters described in Section~\ref{sec:phys_model}.

\section{Physical Forward Model}\label{sec:phys_model}

\subsection{Magnetic configuration and open field-line footpoints}
\label{sec:phys_field}

The near-surface magnetic field is modeled as the superposition of an offset magnetic dipole and an offset axisymmetric ($m=0$) quadrupole, following \citetalias{Kalapotharakos2021}. The dipole component is characterized by five parameters: its Cartesian offset from the stellar center, $(x_D,y_D,z_D)$, inclination angle $\alpha_D$, and azimuthal direction $\phi_D$. The quadrupole component is similarly parameterized by $(x_Q,y_Q,z_Q,\alpha_Q,\phi_Q)$, with an additional parameter $B_Q/B_D$ specifying the quadrupole-to-dipole field-strength ratio at the stellar surface. This defines the 11-dimensional magnetic-field parameter space explored by the MCMC.

For each magnetic-field parameter set, the open-field-line footpoints on the stellar surface are determined by integrating magnetic field lines from a $601\times601$ uniform-solid-angle surface grid using an adaptive Runge--Kutta method. A field line is classified as open if its maximum cylindrical radius exceeds the light-cylinder radius $R_{\rm LC}$ during the integration. The resulting open-field-line footpoints define the emitting surface regions used in the light-curve calculation.

The emitting regions are separated according to magnetic polarity. Without loss of generality, surface elements belonging to the northern magnetic polarity, defined by the positive sign of the radial magnetic field at the surface, are assigned the higher-temperature emission prescription, with luminosity weight $T_w$ and beaming exponent $c_1$, while elements belonging to the southern magnetic polarity are assigned unit luminosity weight and beaming exponent $c_2$.

\subsection{Ray tracing and light-curve construction}
\label{sec:phys_raytrace}

Photon trajectories connecting a distant observer plane to the stellar surface are computed with the \textsc{GIKS} ray-tracing code, using the implementation described by \citet{Lechien2026} and the modeling setup of \citetalias{Kalapotharakos2021}. Rather than integrating photon paths for each magnetic-field configuration, we use a precomputed library of $\sim$$2.5\times10^{6}$ photon trajectories generated for the adopted mass, radius, and observer inclination of \psrseven\ listed in Section~\ref{sec:emission_calibration}. Each trajectory record encodes the surface-intersection coordinates, the arrival time relative to rotational phase, and the local emission angle $\alpha_k$ entering Equation~\ref{eq:photon_weight}.

For a given magnetic-field parameter set, the model light curve in phase bin $i$ is computed as
\begin{equation}
M_i =
A_{\rm cal}
\sum_{k\in\mathcal{R}_i}
w_k\,\mathcal{W}_k ,
\label{eq:lc}
\end{equation}
where $\mathcal{R}_i$ denotes the set of photon trajectories whose arrival phases fall in bin $i$ and whose surface intersections lie within one of the open-field-line footpoints. The factor $w_k$ is the photon-library weight, while $\mathcal{W}_k$ is the additional luminosity weight defined in Equation~\ref{eq:photon_weight}. The normalization $A_{\rm cal}$ is determined by the calibration to the \citet{Dittmann2024} reference light curve.

The result is a 32-bin bolometric model light curve corresponding to the chosen magnetic-field configuration. These light curves are the physical forward-model outputs used to generate the neural-network training databases and to define the surrogate-accelerated model family compared with the observed {\it NICER} bolometric profile.


\section{Neural-Network Surrogate Model}\label{sec:nn}
The physical forward model described in Section~\ref{sec:phys_model} is too computationally expensive to evaluate directly for MCMC exploration at the depth required in this work. Each production MCMC run comprises $1.6\times10^{10}$ likelihood evaluations, completing in $\sim$48 hours on 4000 CPU cores with the neural-network surrogate; at the $\gtrsim\!400\times$ slower evaluation rate of the physical model, an equivalent run would require approximately $400\times48\,\mathrm{hr}\approx 19{,}200\,\mathrm{hr}\approx 2\,\mathrm{years}$ on the same system, making direct physical-model MCMC exploration entirely infeasible. We therefore train neural-network surrogates to emulate the 32-bin bolometric light curves produced by the physical model, following the approach of \citetalias{Olmschenk2025}, and use them for rapid light-curve generation during MCMC. This section describes the training database, network architecture, training procedure, and validation of the surrogate models.

\subsection{Training-database generation}\label{sec:nn_db}
For each of the two temperature-weight prescriptions considered in this work, $T_w=1.31$ and $T_w=1.41$, we generated a synthetic training database using the physical forward model. The offset coordinates were drawn uniformly from the prior:
$
{x_D,y_D,z_D,x_Q,y_Q,z_Q}\sim U(-0.5r_\ast,0.5r_\ast),
$
with the additional constraints that the radial offsets of the dipole and quadrupole moments from the stellar center satisfy $(x_D^2+y_D^2+z_D^2)^{1/2}\leq0.7r_\ast$ and $(x_Q^2+y_Q^2+z_Q^2)^{1/2}\leq0.7r_\ast$, respectively, as in \citetalias{Kalapotharakos2021}.
The angular parameters were drawn as
${\phi_D,\phi_Q}\sim U(0,2\pi)$ and
${\alpha_D,\alpha_Q}\sim U(0,\pi)$, and the quadrupole-to-dipole field-strength ratio was drawn as
$B_Q/B_D\sim U(0.5,11)$.
For each of $\sim 1.6\times10^6$ independent parameter sets, the physical model was used to compute a 32-phase-bin light curve. We then generated 31 additional samples by exploiting the rotational symmetry of the problem: the dipole and quadrupole offset vectors and azimuthal moment angles were rotated about the spin axis in increments of $2\pi/32$, and the corresponding light curve was shifted by the same number of phase bins. This augmentation increases the database size by a factor of 32 without additional physical-model evaluations. The final database contains $5.12\times10^7$ parameter--light-curve pairs. The first $10^5$ samples are reserved for testing, the next $10^5$ for validation, and the remaining samples for training, following the \texttt{haplo} convention used by \citetalias{Olmschenk2025}. Since $10^5$ is an integer multiple of 32, all rotationally augmented copies are kept within the same data split. 

\subsection{Network architecture}\label{sec:nn_arch}
We adopt the open-source \texttt{haplo} neural-network framework developed for \citetalias{Olmschenk2025}; the public source repository is available at \url{https://github.com/golmschenk/haplo} and the software release is archived on Zenodo \citep{Olmschenk2025haplo}. We use the same architecture introduced there (referred to as \texttt{Cura} in the \texttt{haplo} codebase), modifying only a single upsampling value so that the output dimension matches the $N_{\rm bin}=32$ phase bins of the \psrseven\ pulse profile. The network maps the 11 magnetic-field parameters $(x_D,y_D,z_D,\alpha_D,\phi_D,x_Q,y_Q,z_Q,\alpha_Q,\phi_Q,B_Q/B_D)$ to a phase-resolved bolometric X-ray light curve. The architecture is a one-dimensional transposed-convolutional ResNet-style network \citep{He2016} consisting of two initial one-dimensional convolutional dense layers, 19 residual generation blocks, and a final one-dimensional convolutional layer, with the six residual upsampling groups using channel widths $[512,512,1024,1024,2048,2048]$. Our only modification for \psrseven\ is to the final upsampling group, which we set to unity rather than two, so that the network outputs 32 phase bins matching the $N_{\rm bin}=32$ binning of the \citet{Dittmann2024} bolometric profile (Section~\ref{sec:target_data}), rather than the 64 bins used for the PSR~J0030$+$0451 profile in \citetalias{Olmschenk2025}; all other residual blocks, channel widths, activation functions, and hyperparameters are retained. The resulting 32-bin surrogate, which we call \texttt{CuraJ0740}, has 60 trainable layers and $\sim 2\times10^{7}$ trainable parameters. As noted in \citetalias{Olmschenk2025}, the architecture contains no structure explicitly designed for this specific physical model and can be trained to emulate other physical models.

\subsection{Training procedure}\label{sec:nn_train}
We trained independent \texttt{CuraJ0740} models for the two temperature-weight prescriptions, $T_w=1.31$ and $T_w=1.41$, using the corresponding synthetic light-curve databases. Following \citetalias{Olmschenk2025}, the training loss is the median-normalized squared error (MdNSE),
\begin{equation}
{\rm MdNSE} =
\frac{\sum_i (y_i-\hat{y}_i)^2}
{\left[{\rm median}(\hat{y})\right]^2},
\label{eq:mdnse}
\end{equation}
where $y_i$ is the network prediction in phase bin $i$ and $\hat{y}_i$ is the corresponding physical-model value, and the median in the denominator is taken over the phase bins of the target light curve. This normalization prevents the loss from being dominated by high-amplitude light curves and makes the metric sensitive to relative reconstruction accuracy across the prior volume. As in \citetalias{Olmschenk2025}, the optimized quantity is the logarithm of this metric. 

Input parameters and output light curves are standardized using the mean and standard deviation computed from the J0740 training database. During training, the predicted light curves are transformed back to physical units before the loss is evaluated. Training uses the AdamW optimizer
\citep{LoshchilovHutter2019} with learning rate $10^{-4}$, weight decay $10^{-4}$, optimizer $\epsilon=10^{-7}$, and batch size 100. All activations are Leaky ReLUs, and trainable weights are initialized with Kaiming uniform initialization \citep{He2015}. For each temperature-weight prescription, the checkpoint with the lowest validation loss is used for subsequent surrogate-based analysis.

\subsection{Validation}\label{sec:nn_validation}
\begin{figure}[ht]
\centering
\includegraphics[width=0.85\columnwidth]{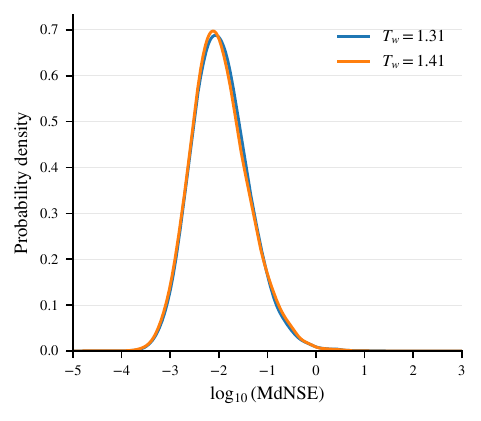}
\caption{%
Test-set accuracy of the two neural-network surrogate models. The curves show the distribution of $\log_{10}(\mathrm{MdNSE})$ for $10^5$ held-out light curves, evaluated separately for the networks trained with $T_w=1.31$ (blue) and $T_w=1.41$ (orange). The two distributions nearly overlap, with median $\log_{10}(\mathrm{MdNSE})\approx -2$, indicating similar surrogate accuracy for the two emission prescriptions.
}
\label{fig:mdnse}
\end{figure}

\begin{figure*}[ht]
\centering
\includegraphics[height=0.85\textheight, keepaspectratio]{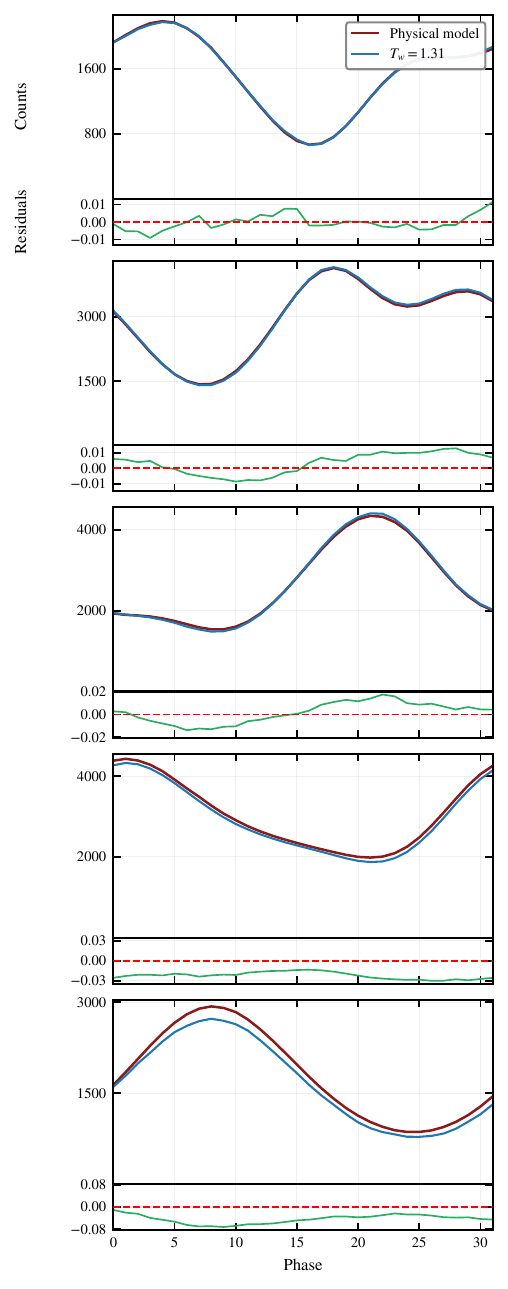}
\quad
\includegraphics[height=0.85\textheight, keepaspectratio]{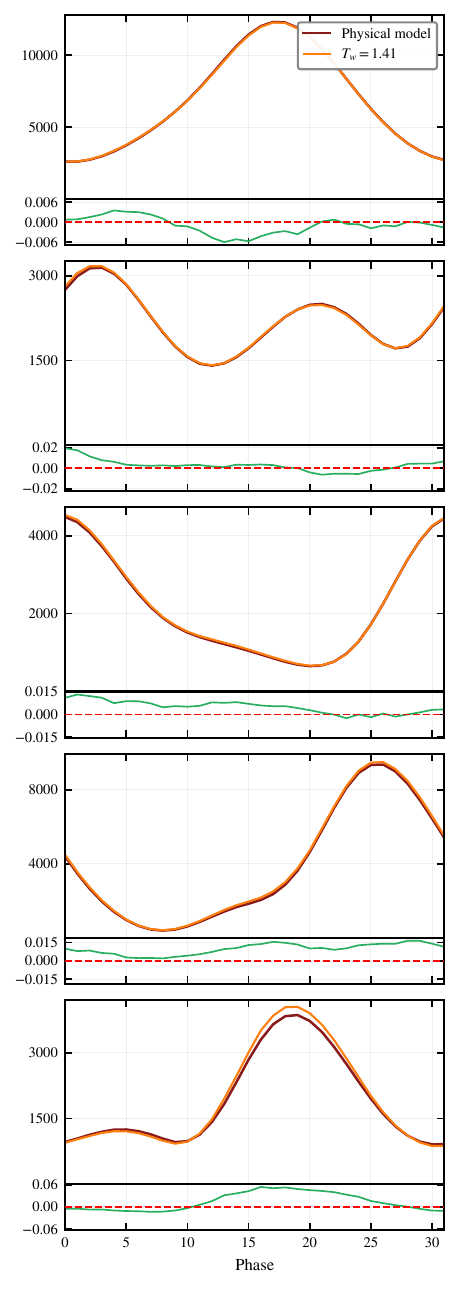}

\caption{Comparison of physical-model light curves (dark red) and neural-network surrogate predictions for $T_w = 1.31$ (left, blue) and $T_w = 1.41$ (right, orange), evaluated on the held-out test set. Each panel shows five representative cases corresponding to the median and the $\pm 1\sigma$ and $\pm 2\sigma$ quantiles of the $\log_{10}(\mathrm{MdNSE})$ distribution, ordered from best to worst (top to bottom) surrogate fidelity. The green curve beneath each comparison shows the residual as a fraction of the maximum amplitude of the physical-model light curve; the horizontal dashed red line marks zero.
}
\label{fig:lc_validation}
\end{figure*}

We validate the trained surrogates on the held-out test set, which is not used during training or checkpoint selection. Figure~\ref{fig:mdnse} shows the distributions of $\log_{10}({\rm MdNSE})$ for the two networks trained with $T_w=1.31$ and $T_w=1.41$, evaluated on $10^5$ test samples drawn
from the prior used to generate the training database. The two distributions overlap closely and have median $\log_{10}({\rm MdNSE})\approx -2$ in both cases. Thus the surrogate accuracy is very similar for the two temperature-weight
prescriptions used to generate the corresponding training database. Figure~\ref{fig:lc_validation} compares physical-model light curves with \texttt{CuraJ0740} predictions for representative held-out examples from each temperature-weight prescription. The residuals are shown as a fraction of the maximum amplitude of the corresponding physical-model light curve. The neural-network predictions closely track the physical-model light curves across the 32 phase bins, and the residuals show no large coherent mismatch in the displayed cases. Together, the MdNSE distributions and representative light-curve comparisons indicate that the trained networks reproduce the physical-model light curves with sufficient fidelity for the subsequent surrogate-based MCMC exploration.

\section{Neural-Network-Accelerated MCMC}\label{sec:mcmc}

The trained neural-network surrogate replaces the physical forward model during MCMC exploration. Following \citetalias{Olmschenk2025}, this substantially reduces the computational cost of each likelihood evaluation. For each of the two calibrated emission prescriptions, $T_w=1.31$ and $T_w=1.41$, we perform an independent MCMC run in the 11-dimensional magnetic-field parameter space described in Section~\ref{sec:phys_field}. All source parameters and emission-calibration quantities discussed in Section \ref{sec:emission_calibration} are held fixed.

For a proposed magnetic-field parameter set, the log-likelihood is
\begin{equation}
\ln\mathcal{L} = -\frac{1}{2}\sum_{i=1}^{32}
\left[\frac{(D_i - M_i)^2}{\sigma_i^2}
+ \ln(2\pi\sigma_i^2)\right],
\label{eq:loglike}
\end{equation}
where $D_i$ is the background-subtracted observed count in phase bin $i$, $M_i$ is the surrogate-predicted model count for the same bin, and $\sigma_i$ is given by Equation~\ref{eq:sigma}.

We use the affine-invariant ensemble sampler of \citet{GoodmanWeare2010}, implemented as two interleaved sub-chains per CPU core following \citetalias{Olmschenk2025}. Each of the 4000 CPU cores evolves two independent walkers. New positions are proposed using the stretch-move, with the complementary sub-ensemble as the reference ensemble. Initial walker positions are drawn uniformly from the prior using independent random seeds on each core. Each run comprises $2\times10^6$ iterations per walker, yielding a total of $ 1.6\times10^{10}$ likelihood evaluations (4000 CPU cores $\times$ 2 walkers per core $\times$ $2\times10^6$ iterations per walker).

Storing and analyzing all $1.6\times10^{10}$ samples is impractical, so we retain a uniform random subsample, keeping each sample independently with probability $p\approx6.25\times10^{-4}$ while preserving its iteration index. This yields $\sim10^{7}$ samples that uniformly represent the full chain, from which all posterior results below are derived. Burn-in is removed after subsampling (Section~\ref{sec:res_conv}); unless stated otherwise, posterior distributions use the last $10\%$ of the post-burn-in chain, corresponding to $\sim9.51\times10^{5}$ subsampled points.

\section{Results}\label{sec:results}

\subsection{MCMC convergence}\label{sec:res_conv}

\begin{figure*}[ht]
  \centering
    \raisebox{0.0cm}{
  \includegraphics[width=0.51\textwidth]{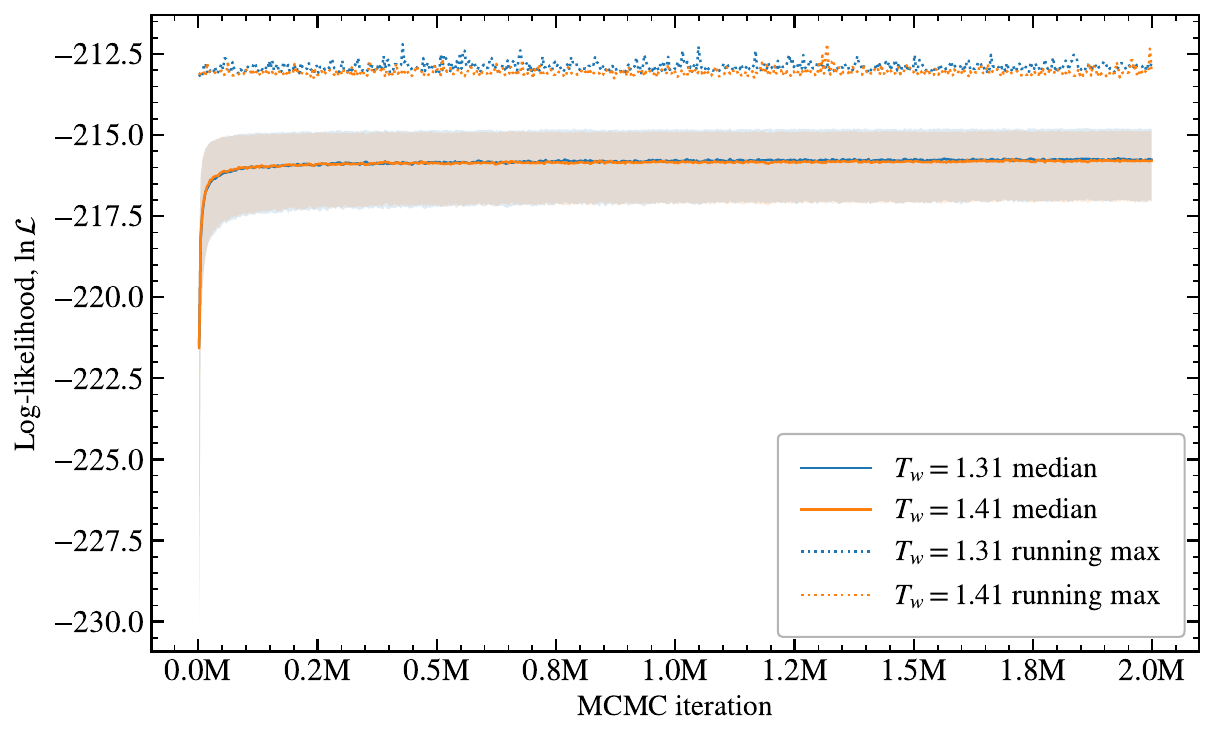}}
  \hfill
  \raisebox{-0.05cm}{%
  \includegraphics[width=0.45\textwidth]{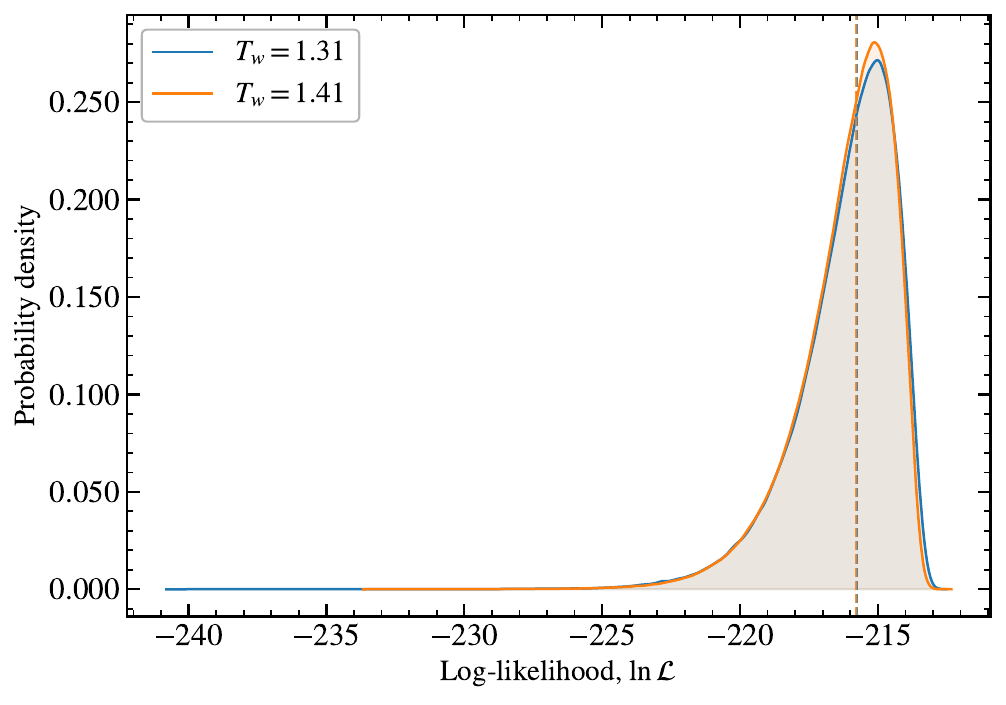}
}
  \caption{%
    Log-likelihood convergence of the neural-network-accelerated-MCMC chains for \psrseven\
    under $T_w = 1.31$ (blue) and $T_w = 1.41$ (orange).
    \textbf{\textit{Left}}: Full $2\times10^6$-iteration run. Solid lines show the binned median $\ln\mathcal{L}$ across all chains; shaded bands indicate the inter-quartile range (IQR, 25th--75th percentile); dotted lines show the running maximum.
    \textbf{\textit{Right}}: A comparison of distributions of the log-likelihood over the last 10\% of the post-burn-in chain. Dashed vertical lines mark the median $\ln\mathcal{L}$ for each run. The close overlap of the two distributions, with both medians at $\approx -215.75$, indicates that the two runs sample regions of comparable likelihood.
  }
  \label{fig:loglik_convergence}
\end{figure*}

Figure~\ref{fig:loglik_convergence} (left) shows the evolution of the median $\ln\mathcal{L}$ and interquartile range across all chains over the full run. In both cases, the likelihood improves rapidly during the burn-in phase before stabilizing; we adopt a conservative burn-in cutoff at $10^5$ iterations, corresponding to 5\% of the total chain length, and restrict all subsequent analysis to the post-burn-in samples. The highest log-likelihood values found are $\ln\mathcal{L}_{\rm max} = -212.52$ for $T_w = 1.31$ and $\ln\mathcal{L}_{\rm max} = -212.33$ for $T_w = 1.41$.

Figure~\ref{fig:loglik_convergence} (right) shows the kernel density estimate (KDE) of the $\ln\mathcal{L}$ distribution over the last 10\% of the post-burn-in samples for each run. The two distributions overlap closely and peak at nearly identical values, with both medians at $\approx -215.75$ (dashed vertical lines). This similarity indicates that, after burn-in, the two calibrated emission prescriptions sample regions of the magnetic-field parameter space with very similar likelihood values. 

\subsection{Posterior distributions}\label{sec:res_posterior}

\begin{figure*}[ht]
  \centering
  \includegraphics[width=\textwidth]{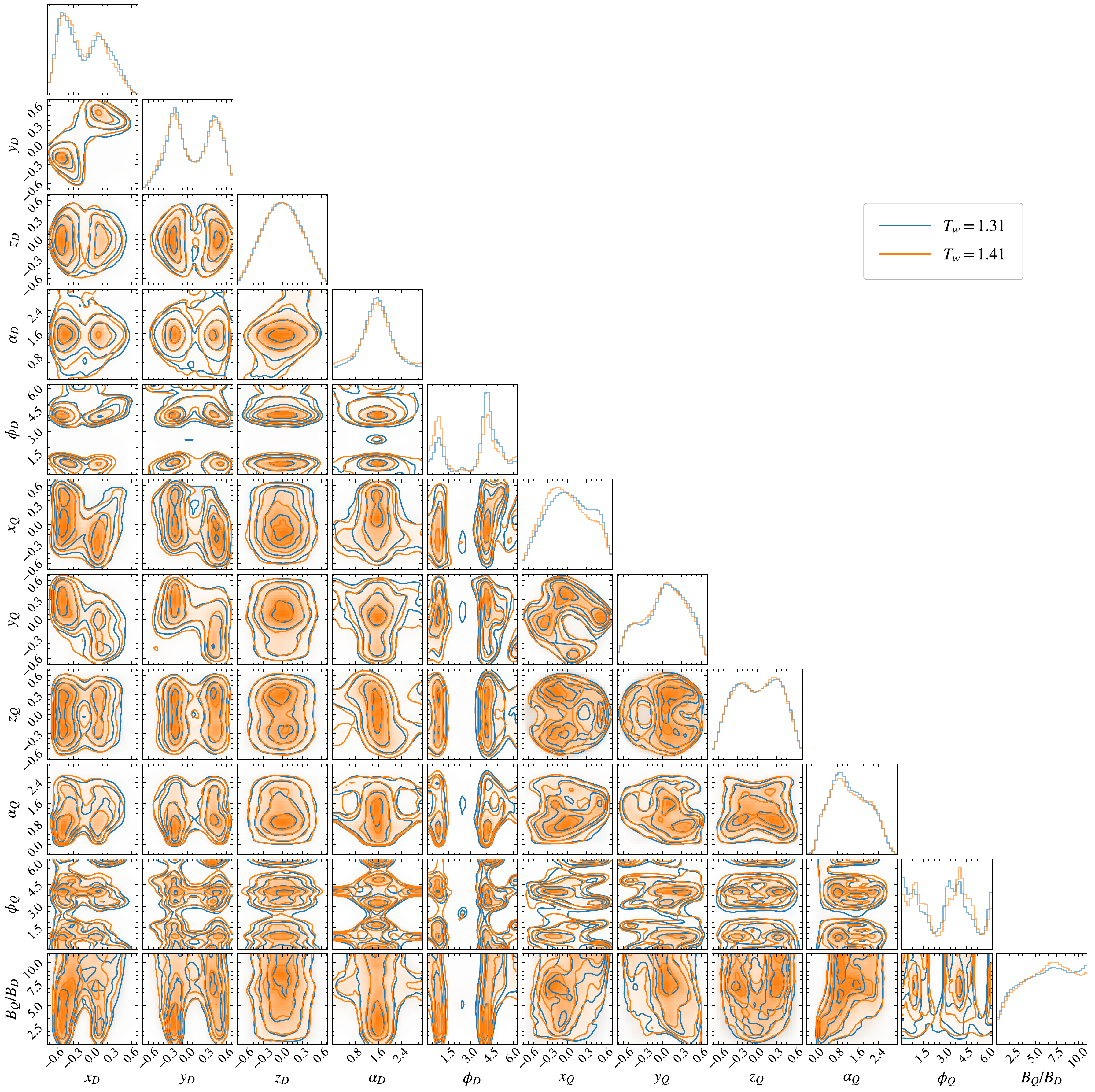}
  \caption{%
    Joint posterior distribution of the 11 magnetic-field parameters for \psrseven, overlaying results from $T_w = 1.31$ (blue) and $T_w = 1.41$ (orange). Each posterior is constructed from the last 10\% of the respective post-burn-in chain ($\sim$$9.51\times10^5$ samples, each subsampled to $1\times10^5$ for display). Contours in the off-diagonal panels enclose the 68\% and 95\% credible regions. Diagonal panels show the marginalized 1-D posterior for each parameter. The two distributions are consistent across all parameters, demonstrating that the inferred magnetic-field geometry is robust to the adopted value of $T_w$ within the range explored.
  }
  \label{fig:corner_overlay}

\end{figure*}

Figure~\ref{fig:corner_overlay} shows the joint posterior distribution of the 11 magnetic-field parameters, overlaying the results from the two temperature-weight runs, $T_w = 1.31$ (blue) and $T_w = 1.41$ (orange), each constructed from the last 10\% of the respective post-burn-in chain. We verified that posterior distributions constructed from earlier post-burn-in segments, after the likelihood evolution has stabilized, are very similar to those shown here. 

The two temperature-weight prescriptions produce nearly indistinguishable posterior structure across the 11-dimensional magnetic-field parameter space. This indicates that, within the two calibrated emission prescriptions considered here, the inferred magnetic-field structure is not appreciably sensitive to the adopted value of $T_w$. 

The posterior distributions are broad in most parameters, reflecting the limited constraining power of the background-dominated bolometric pulse profile of \psrseven\ (Section~\ref{sec:target_data}). Several parameters exhibit multimodal structure; this is a generic feature of the offset dipole-plus-quadrupole parameterization, which
admits multiple field configurations producing similar hotspot footpoints and hence similar pulse profiles, as previously noted for PSR~J0030$+$0451 \citepalias{Kalapotharakos2021,Olmschenk2025}.
The quadrupole strength $B_Q/B_D$ is broadly distributed and extends toward the upper prior boundary of 11. This indicates that substantial quadrupolar contributions are allowed within the adopted parametrization, but the upper-boundary behavior should be interpreted with caution.


\begin{table}[ht]
\centering
\caption{Marginalized posterior medians and 68\% credible intervals (16th--84th percentiles) for the 11 magnetic-field parameters of \psrseven, for both temperature-weight prescriptions. Offset coordinates $(x_D, y_D, z_D, x_Q, y_Q, z_Q)$ are in units of $r_\ast$; inclination angles $(\alpha_D, \alpha_Q)$ and azimuthal angles $(\phi_D, \phi_Q)$ are in radians. The broad intervals reflect the low signal-to-noise regime of \psrseven\ (Section~\ref{sec:target_data}); for multimodal parameters (notably $\phi_D$ and $\phi_Q$) the quoted median and intervals summarize the full marginal distribution and should be interpreted alongside Figure~\ref{fig:corner_overlay}.}
\label{tab:posterior}

\begin{tabular}{lcc}
\hline\hline
Parameter & $T_w = 1.31$ & $T_w = 1.41$ \\
\hline
    $x_D$        & $-0.161^{+0.416}_{-0.326}$ & $-0.189^{+0.392}_{-0.289}$ \\[2pt]
    $y_D$        & $+0.068^{+0.401}_{-0.339}$ & $+0.037^{+0.437}_{-0.333}$ \\[2pt]
    $z_D$        & $-0.008^{+0.299}_{-0.301}$ & $-0.017^{+0.306}_{-0.300}$ \\[2pt]
    $\alpha_D$   & $+1.563^{+0.609}_{-0.602}$ & $+1.550^{+0.675}_{-0.694}$ \\[2pt]
    $\phi_D$     & $+4.089^{+0.858}_{-3.194}$ & $+3.883^{+1.075}_{-3.195}$ \\[2pt]
    $x_Q$        & $+0.028^{+0.389}_{-0.356}$ & $-0.034^{+0.408}_{-0.326}$ \\[2pt]
    $y_Q$        & $+0.068^{+0.327}_{-0.425}$ & $+0.056^{+0.326}_{-0.406}$ \\[2pt]
    $z_Q$        & $+0.016^{+0.371}_{-0.389}$ & $+0.017^{+0.362}_{-0.389}$ \\[2pt]
    $\alpha_Q$   & $+1.204^{+0.892}_{-0.677}$ & $+1.236^{+0.890}_{-0.701}$ \\[2pt]
    $\phi_Q$     & $+3.206^{+1.582}_{-2.497}$ & $+3.298^{+1.486}_{-2.450}$ \\[2pt]
    $B_Q/B_D$    & $+6.380^{+3.130}_{-3.505}$ & $+6.430^{+2.966}_{-3.413}$ \\[2pt]
\hline
\end{tabular}
\end{table}

  \subsection{Light-curve ensemble}\label{sec:res_lc}
As a posterior-predictive check at the level of the observable, we draw $10^{4}$ parameter samples uniformly from the late-time post-burn-in chain 
and pass each sample through \texttt{CuraJ0740} in inference mode to obtain a predicted light curve. Unless otherwise stated, we use samples from the final 10\% of the chain; using a broader late-time interval, iterations $1.5\!\times\!10^{6}$--$2\!\times\!10^{6}$, gives statistically indistinguishable results. The ensemble of predicted light curves defines, at each phase bin, a posterior distribution of model counts.

\begin{figure}[ht]
  \centering
  \includegraphics[width=\columnwidth]{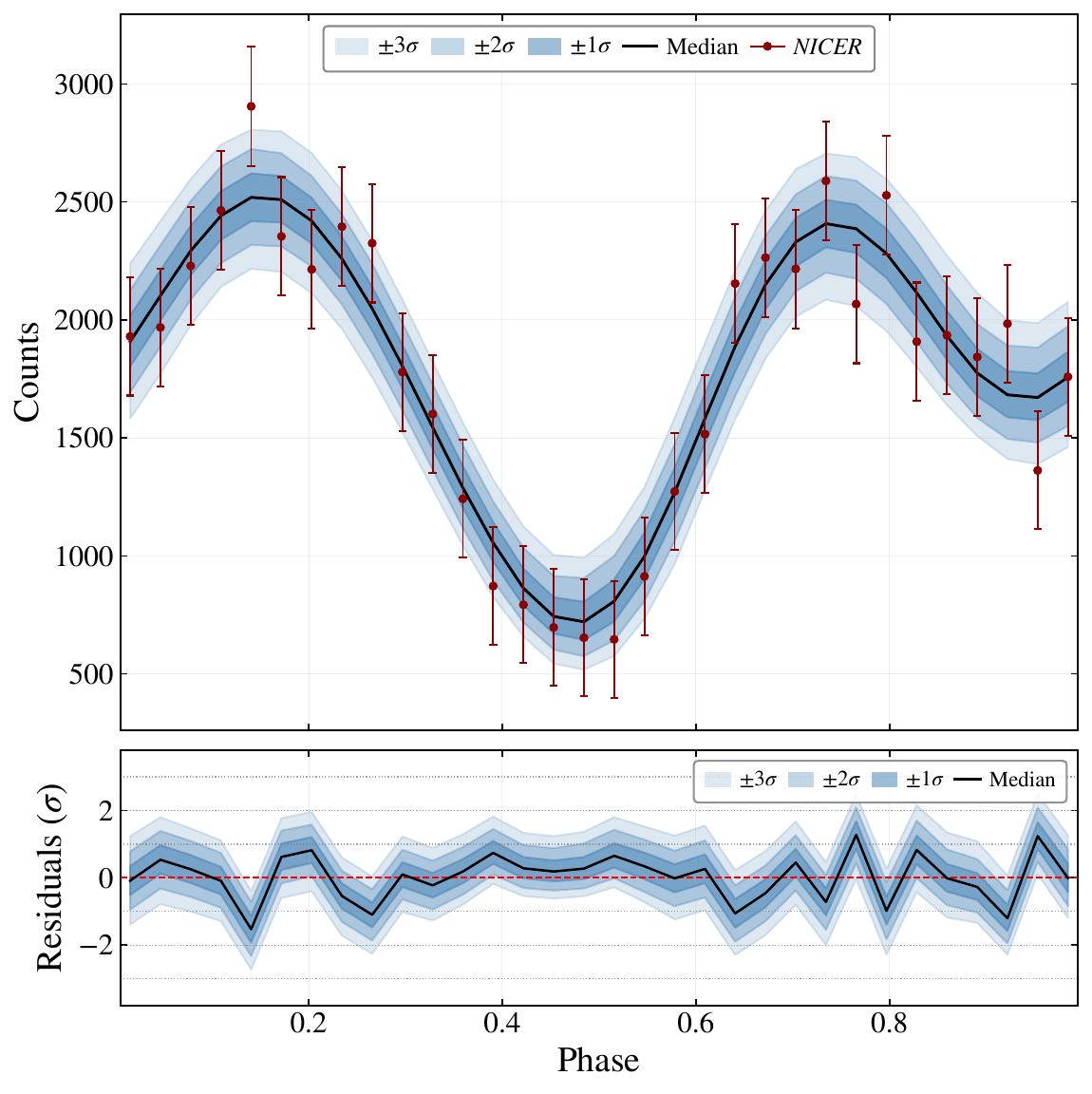}
  \caption{Posterior predictive light-curve ensemble for PSR J0740+6620 using $10^{4}$ parameter samples drawn from iterations $1.5\!\times\!10^{6}$--$2\!\times\!10^{6}$ of the post-burn-in chain for the  $T_w=1.41$ run. {\it Top}: background-subtracted NICER observed counts (red points with $1\sigma$ error bars), overlaid with the $\pm 1\sigma$ (dark), $\pm 2\sigma$ (medium), and $\pm 3\sigma$ (light) contours of the predicted distribution and the median model (black line). {\it Bottom}: normalized residuals $(M_{i}-D_{i})/\sigma_{i}$ for the ensemble, with the same contour levels. The median residual (black) lies near zero across all 32 phase bins.}
  \label{fig:lightcurve_ensemble}
\end{figure}

The top panel of Figure~\ref{fig:lightcurve_ensemble} shows the observed NICER pulsed counts, after background subtraction, with the $\sigma_{i}$ uncertainties of Equation~\ref{eq:sigma}. These are overlaid with the median model and the $\pm 1\sigma$, $\pm 2\sigma$, and $\pm 3\sigma$ contours of the light-curve ensemble. The lower panel shows the normalized residuals $(M_{i}-D_{i})/\sigma_{i}$ for the same ensemble.

The posterior light-curve ensemble reproduces the main features of the observed profile. The median residual remains close to zero across the 32 phase bins, indicating no obvious phase-dependent bias. 
The ensemble bands are comparatively wide, with typical $\pm 1\sigma$ widths of order the observational uncertainty, $\sigma_{i}\sim 250$ counts. This width is dominated by the genuine breadth of the posterior in field-configuration space, not by the neural-network surrogate error: the MdNSE values for samples drawn from the posterior region  are well below the values that would correspond to $1\sigma$ widths of this magnitude.


\subsection{Hotspot geometry}\label{sec:res_hotspots}
To characterize the surface emission geometry implied by the posterior, we recompute the open-field-line footpoints, i.e., hotspots, with the physical forward model for $10^4$ post-burn-in posterior samples. For each sample, the emitting regions are determined self-consistently from the corresponding offset dipole-plus-quadrupole vacuum field. We then accumulate the resulting hotspot maps over the posterior sample, so that the value at each surface location gives the fraction of posterior samples in which that location belongs to an emitting open-field-line footpoint.

\begin{figure*}[ht]
  \centering
  \includegraphics[width=\textwidth]{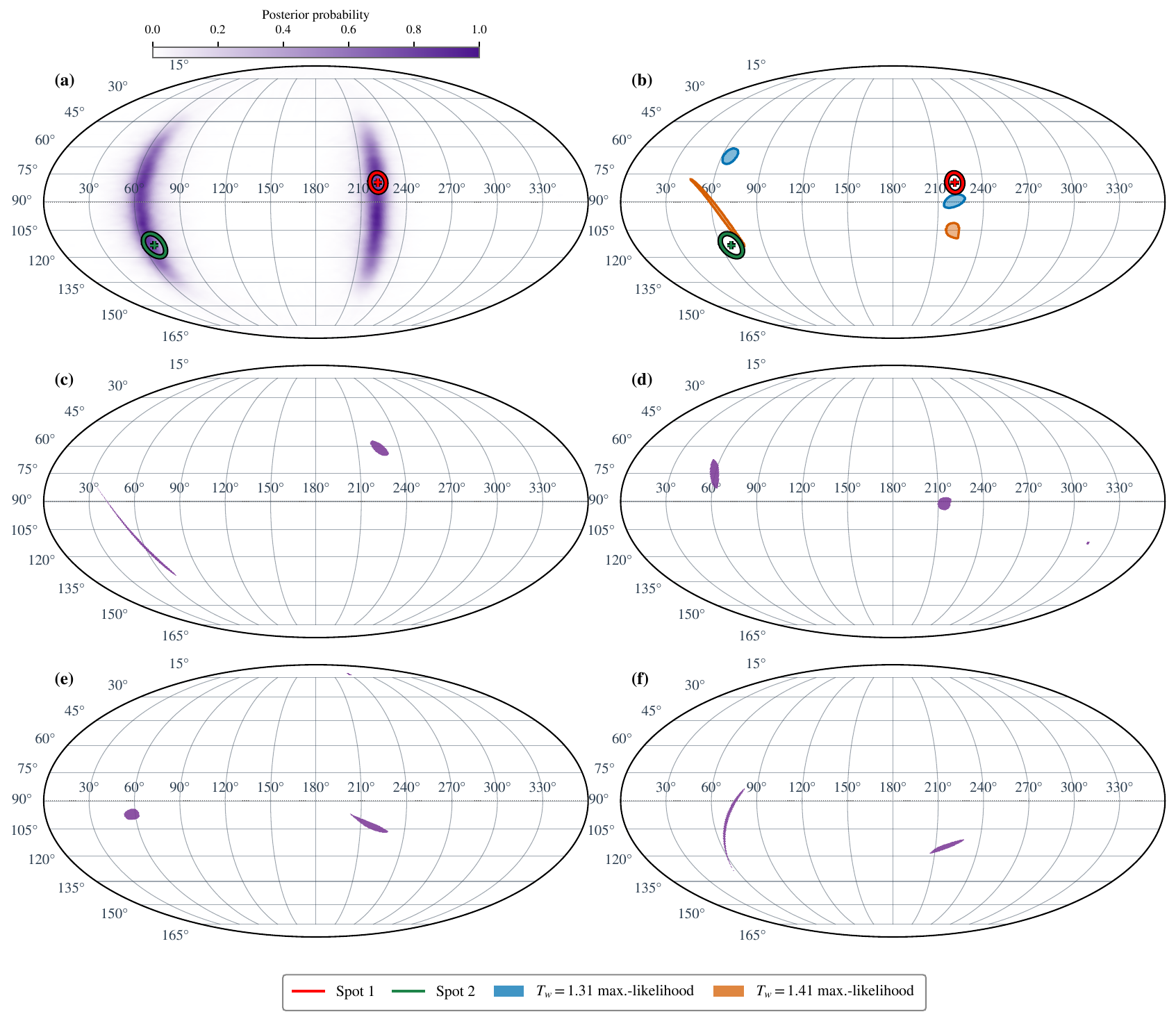}
  \caption{%
    Surface emission geometry of \psrseven\ inferred from the neural-network-accelerated MCMC posterior, shown on Mollweide projections of the stellar surface.
    \textbf{(a)} Accumulated hotspot density map constructed from $10^4$ post-burn-in posterior samples.
    The color scale indicates the fraction of posterior samples for which each surface element belongs to the open-field-line emission region. Solid red and green curves show the maximum-likelihood circular hotspot boundaries of \citet{Dittmann2024} for the primary
    (Spot~1) and secondary (Spot~2) emission regions, respectively. \textbf{(b)} Maximum-likelihood open-field-line footpoints from the $T_w = 1.31$ (blue) and $T_w = 1.41$ (orange) runs, overlaid with the same \citet{Dittmann2024} hotspots. The footpoints emerge self-consistently from the offset dipole-plus-quadrupole vacuum field and are not constrained
    to be circular.
    \textbf{(c)--(f)} Representative individual open-field-line footpoints drawn from the $T_w = 1.31$ posterior, selected to illustrate the range of magnetic-field configurations consistent with the \nicer\ data. The variety of footpoint morphologies --- from compact near-equatorial regions to extended arc-shaped configurations --- reflects the breadth of the posterior and the multimodal structure discussed throughout this section.
  }
  \label{fig:hotspots}
\end{figure*}

Figure~\ref{fig:hotspots} shows the resulting surface-emission geometry in Mollweide projection. Panel~(a) presents the accumulated posterior hotspot-density map. The azimuthal locations of the emitting regions are relatively well constrained, as expected from the requirement that the model reproduce the observed pulse phases. In contrast, the allowed emission regions remain extended in latitude and surface extent, and the posterior density is not tightly concentrated around the maximum-likelihood circular hotspots of \citet{Dittmann2024}. The \citet{Dittmann2024} hotspot boundaries are therefore shown only as a reference comparison, not as imposed targets for the magnetic-field-generated footpoints. 

Panel~(b) shows the maximum-likelihood open-field-line footpoints obtained for the $T_w=1.31$ (blue) and $T_w=1.41$ (orange) runs. The two cases produce broadly similar emission locations, consistent with the weak dependence of the magnetic-field posterior on the adopted temperature-weight prescription. The footpoints are not constrained to be circular and can differ appreciably in shape from the reference circular hotspots, as expected for emission regions generated by multipolar magnetic fields.

Panels~(c)--(f) show representative individual footpoints drawn from the $T_w=1.31$ posterior. These examples illustrate the range of geometries allowed by the data, including compact regions and extended arc-like structures. The diversity of the individual footpoints reflects the broad and multimodal posterior discussed above: distinct magnetic-field configurations can produce light curves that are statistically similar while corresponding to different detailed surface-emission morphologies.


\section{Discussion}\label{sec:discussion}

The goal of this work is to ask what magnetic-field structures can reproduce the bolometric \nicer\ pulse profile of \psrseven\ when the emitting regions are required to arise from magnetic open-field-line footpoints rather than from prescribed geometric hotspots. The neural-network-accelerated inference shows that an offset dipole-plus-quadrupole static-vacuum field can reproduce the observed double-peaked light curve, while allowing a broad and multimodal family of magnetic-field configurations. In this section, we discuss the robustness of the inference, the role of offsets and quadrupolar structure, the relation of the inferred surface geometry to geometric hotspot models, and the main limitations of the present approach.

\subsection{Robustness of the inference and the role of offsets}
\label{sec:disc_offsets}

The comparison between the $T_w=1.31$ and $T_w=1.41$ runs provides an internal check on the sensitivity of the inference to the relative weighting of the two emitting regions. The parameter $T_w$ encodes the relative bolometric weight associated with the effective-temperature difference between the hotspots. The posterior distributions in Figure~\ref{fig:corner_overlay}, the posterior-predictive light-curve ensembles in Figure~\ref{fig:lightcurve_ensemble}, and the maximum-likelihood surface footpoints in Figure~\ref{fig:hotspots} show only weak differences between the two prescriptions. Thus, within the calibrated range explored here, changing the relative bolometric hotspot weighting does not materially alter the inferred magnetic-field configuration.

\begin{figure*}[ht]
  \centering
  \includegraphics[width=0.7\textwidth]{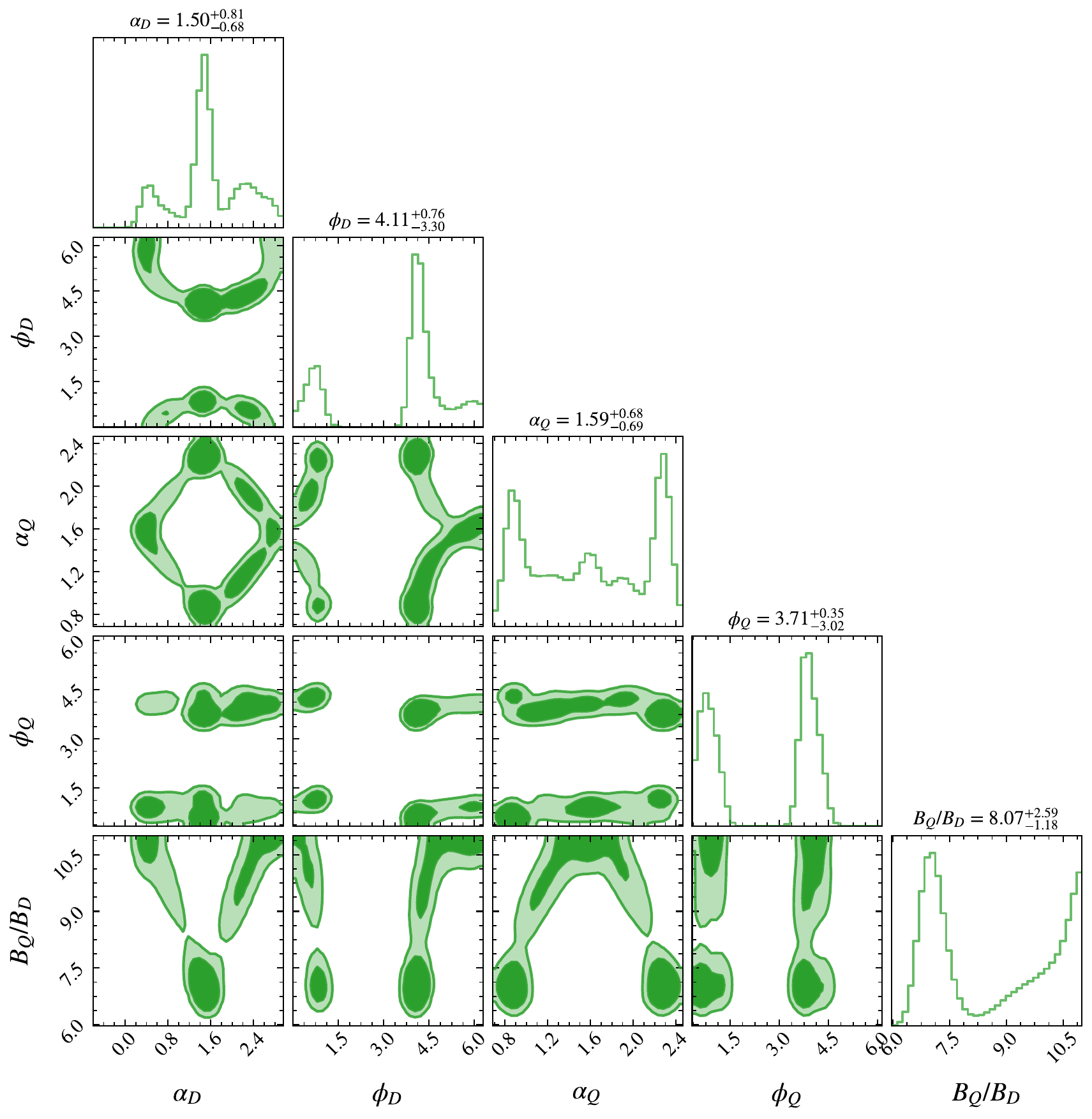}
  \caption{%
    Posterior distribution for the centered, five-parameter magnetic-field model. In this run, both magnetic moments are fixed at the stellar center, so $x_D=y_D=z_D=x_Q=y_Q=z_Q=0$. The remaining parameters are the dipole and quadrupole orientations and the strength ratio $B_Q/B_D$. The posterior is constructed from the final 10\% of the post-burn-in samples for the $T_w=1.41$ run. This model tests whether a centered dipole plus a centered axisymmetric quadrupole can reproduce the observed pulse profile. We find that it cannot match the observed profile as closely as the offset model (median $\chi^2_r \approx 1.50$ versus $\approx 0.92$--$0.93$; Figure~\ref{fig:red_chi_sq}).
  }
  \label{fig:corner_zero_offset}
\end{figure*}

Because the full 11-parameter posterior is broad, we also tested whether the data require spatial offsets of the magnetic moments within the adopted field family. For this purpose, we performed an additional MCMC run in which both the dipole and quadrupole moments were fixed at the stellar center. This reduces the magnetic-field model to five parameters, corresponding to the two moment orientations and the quadrupole-to-dipole strength ratio. The corresponding posterior distributions are shown in Figure~\ref{fig:corner_zero_offset}. This run explores the possibility that the observed profile can be explained by a centered dipole plus a centered, rotated, axisymmetric quadrupole.

\begin{figure}[ht]
  \centering
  \includegraphics[width=\columnwidth]{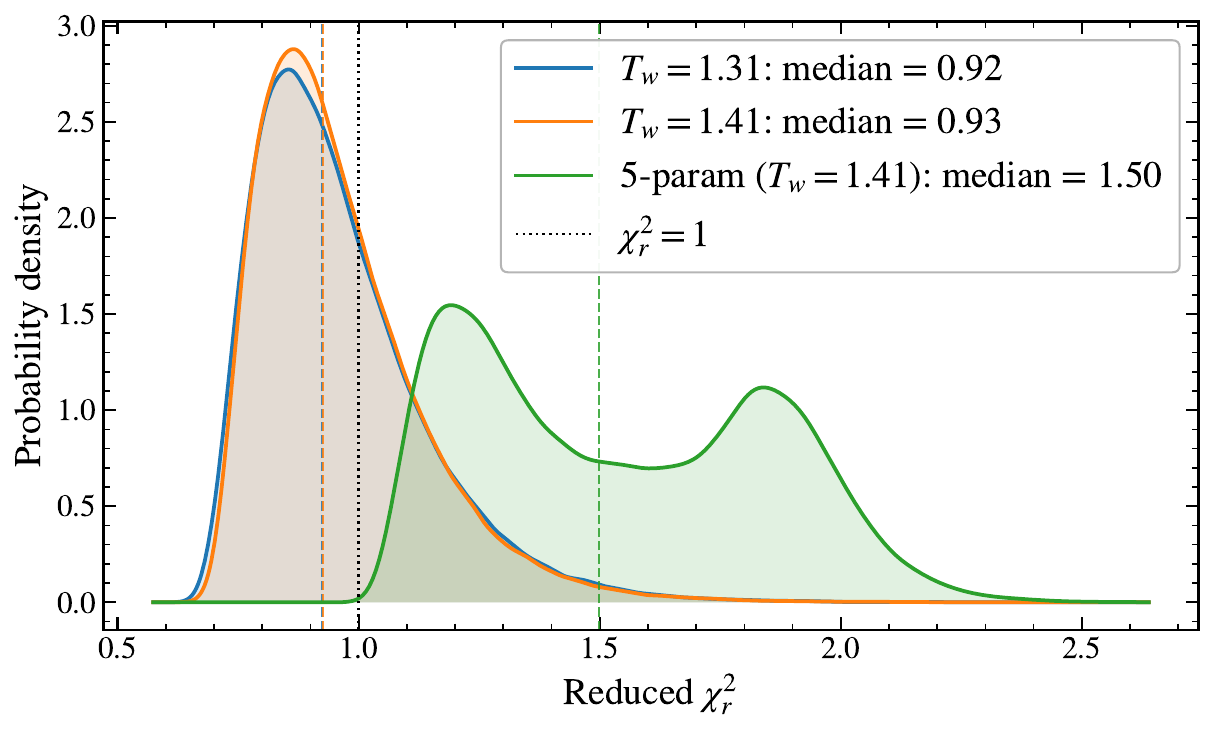}
  \caption{%
    Fit quality for the offset and zero-offset magnetic-field models. The curves show the post-burn-in distributions of reduced chi-squared, $\chi^2_r=(-2\ln\mathcal{L}-C)/{\rm dof}$, where $C=\sum_i\ln(2\pi\sigma_i^2)$. Blue and orange show the full offset dipole-plus-quadrupole model for $T_w=1.31$ and $T_w=1.41$, respectively; green shows the centered five-parameter model for $T_w=1.41$. Dashed lines mark the medians, and the dotted black line marks $\chi^2_r=1$. The zero-offset model gives systematically poorer fits than the offset models within the magnetic-field family considered here.}
    
  \label{fig:red_chi_sq}
\end{figure}

The centered model is disfavored relative to the offset runs within the restricted magnetic-field family considered here. The reduced-$\chi^2$ distributions for the original offset-field runs are centered close to unity, whereas the corresponding distribution for the zero-offset run is shifted to values clearly larger than unity (Figure~\ref{fig:red_chi_sq}). The preference for the offset model is therefore reflected directly in the light-curve fit quality, not only in the structure of the posterior distributions. In this sense, the bolometric \nicer\ profile favors additional surface-field complexity beyond a centered dipole plus centered axisymmetric quadrupole.

This conclusion should be interpreted carefully. The quadrupole component adopted here is the rotated $m=0$ quadrupole only. A fully general quadrupolar magnetic field, including the non-axisymmetric components, would provide additional shape freedom even without spatially offsetting the multipole moments. Therefore, the poorer performance of the zero-offset run does not prove that the physical magnetic moments of \psrseven\ must be displaced from the stellar center. Rather, it shows that, within the restricted dipole-plus-axisymmetric-quadrupole parameterization adopted here, offsets provide the additional flexibility needed to reproduce the observed bolometric pulse profile. More general centered multipolar fields, such as those explored by \citet{Kundu2026}, will be important for determining whether offsets remain necessary in a less restricted magnetic-field basis.

\subsection{Surface geometry and magnetic-moment directions}
\label{sec:disc_surface}

The inferred surface-emission geometry is broad, but not arbitrary. The phases of the observed pulse peaks constrain the azimuthal placement of the emitting regions, while the bolometric profile leaves substantial freedom in their latitude, surface extent, and detailed shape. This behavior is visible in the accumulated hotspot-density map of Figure~\ref{fig:hotspots}: the posterior does not allow emission everywhere, but it also does not collapse onto compact circular regions.

A direct comparison with the geometric-hotspot posterior of \citet{Dittmann2024} is limited by the different model assumptions and by the different data products used in the two analyses. \citet{Dittmann2024} did not present an analogous posterior hotspot-density map. Their posterior distributions for the circular-hotspot center colatitudes, for example $\theta_{c1}$ and $\theta_{c2}$ in their Figure~11, do show extended allowed ranges, although the apparent latitudinal spread is not directly comparable to the accumulated footpoint coverage shown in Figure~\ref{fig:hotspots}. The \citet{Dittmann2024} parameters describe the centers of prescribed circular hotspots, whereas our map shows the surface coverage of open-field-line footpoints generated by an 11-dimensional magnetic-field model. In addition, the present inference uses only the bolometric pulse profile, rather than the full phase-resolved energy-dependent data used in the original pulse-profile analysis. At the same time, our calculation fixes $M$, $R$, and the observer inclination to the highest-likelihood values from \citet{Dittmann2024}, which removes geometric freedom that was included in their full posterior exploration. One might therefore have expected this choice to push the inferred surface geometry closer to their maximum-likelihood configuration. The broad latitudinal and surface-area extent of our hotspot-density map, despite this fixed global geometry, may therefore reflect both the reduced constraining power of the bolometric-only light curve and the additional geometric degeneracies introduced by the magnetic-field parameterization.

\begin{figure*}[ht]
  \centering
  \includegraphics[width=\textwidth]{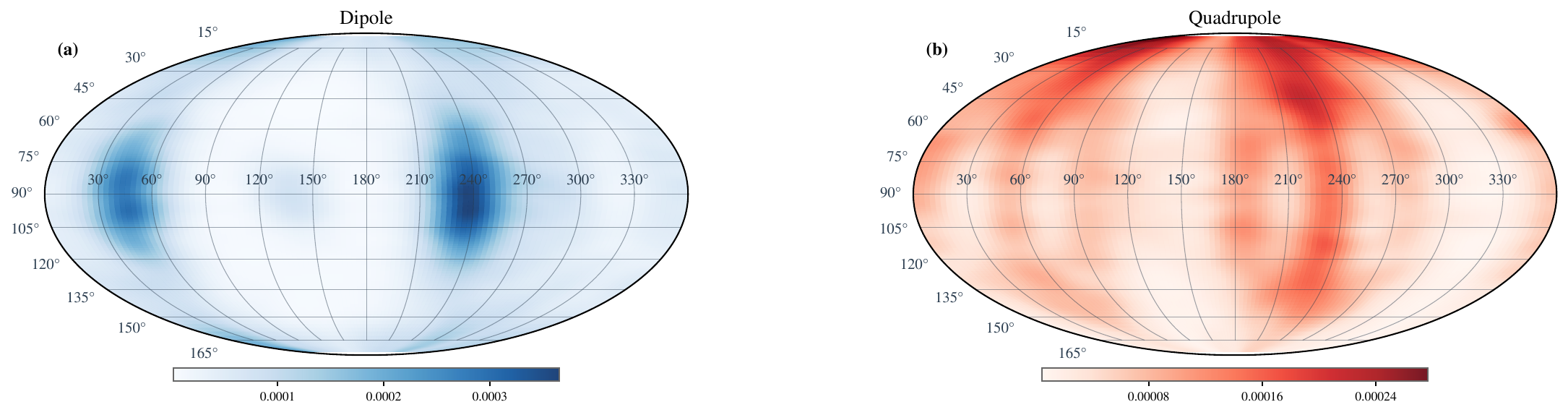}
  \caption{%
    Posterior distribution of magnetic moment directions for \psrseven, constructed from $10^4$ post-burn-in samples of the $T_w = 1.41$ run. \textbf{(a)} Dipole magnetic moment direction, determined by $(\alpha_D, \phi_D)$. \textbf{(b)} Quadrupole magnetic moment direction, determined by $(\alpha_Q, \phi_Q)$. 
  }
  \label{fig:magmoment}
\end{figure*}

The posterior distributions of the magnetic-moment directions provide a complementary view of the inferred geometry. Figure~\ref{fig:magmoment} shows the surface locations intersected by the dipole and quadrupole moment axes for posterior samples from the $T_w=1.41$ run. The dipole moment directions tend to intersect the stellar surface near the regions associated with the inferred hotspots, suggesting that the dipole orientation helps set the approximate emission longitudes. The quadrupole moment directions are more concentrated toward the northern rotational hemisphere. However, the distribution also includes solutions in which the quadrupole moment lies at lower latitudes, closer to the equatorial region and along azimuths similar to those favored by the dipole moment. This indicates that the quadrupole component is not associated with a single unique surface direction. Instead, it provides additional control over the open-field-line topology, allowing the model to reproduce the observed pulse profile through a range of detailed footpoint morphologies.

\subsection{Model limitations and future extensions}
\label{sec:disc_limits}

The magnetic-field model used here is a static-vacuum approximation. It should therefore be viewed as a surface-field parameterization rather than a fully self-consistent pulsar magnetosphere. In a plasma-filled force-free magnetosphere, currents and field-line sweepback can modify the open-field-line boundary, the polar-cap shape, and the mapping between surface footpoints and outer-magnetosphere emission. Previous work on PSR~J0030$+$0451 showed that static-vacuum multipolar fields can nevertheless trace the broad surface-field structures needed to reproduce \nicer\ pulse profiles \citepalias{Kalapotharakos2021}, and ongoing comparisons with force-free models suggest that this remains a useful first approximation \citep{Lechien2026}. Still, the detailed footpoint locations and inferred multipolar parameters should not be interpreted as a unique force-free solution for \psrseven.

A second limitation is that the present inference uses the bolometric light curve rather than the full phase-energy spectrogram. This choice makes the first magnetic-field exploration computationally tractable and allows direct use of the neural-network surrogate, but it discards information carried by the energy dependence of the pulse profile. In addition, the present beaming prescription is a simplified approximation to a full atmosphere model. The hemisphere-dependent exponents partially capture differences in the angular emission pattern, but they do not replace a self-consistent treatment of the spectral and angular emissivity. We also do not explore the possible impact of stellar oblateness on the detailed mapping between surface emission, photon propagation, and the observed pulse profile. Phase-resolved spectroscopy, combined with a more realistic atmosphere model, can help separate temperature, beaming, area, and geometric effects, and may therefore reduce the broad latitudinal and surface-area uncertainties seen in the hotspot-density map. A future extension using the full two-dimensional phase-energy data would provide a more direct comparison with the original pulse-profile analysis of \citet{Dittmann2024}.

Joint X-ray and $\gamma$-ray modeling is especially promising for \psrseven. The X-ray pulse profile constrains the surface-emission geometry, while the $\gamma$-ray light curve probes the global magnetospheric structure and the observer's viewing cut through the high-altitude emission region, associated with the vicinity of the equatorial current sheet in force-free and dissipative models. Combining these data sets can break degeneracies that are difficult to resolve with the bolometric X-ray profile alone, particularly degeneracies between magnetic inclination, observer geometry, multipolar surface structure, and the relative placement of the emitting regions. In this sense, the present work should be regarded as the X-ray surface-field component of a broader multiwavelength inference program.

Finally, the posterior preference for substantial quadrupolar contributions should be interpreted within the adopted prior and model basis. The marginal distribution of $B_Q/B_D$ peaks at large values and extends toward the upper prior boundary, indicating that the current parameter range may not fully capture all allowed quadrupole-dominated configurations. Extending this range would require retraining the neural-network surrogate on a wider prior. More importantly, the present model includes only the rotated axisymmetric quadrupole component. A more complete multipolar expansion, combined with force-free magnetospheric structure and joint X-ray/$\gamma$-ray fitting, will be needed to determine how uniquely the current data require offset moments, strong quadrupolar structure, or both.

\section{Conclusions}\label{sec:conclusions}

We have presented a neural-network-accelerated magnetic-field inference for \psrseven, modeling the \nicer\ bolometric pulse profile with emission regions generated by the open-field-line footpoints of an offset dipole-plus-quadrupole static-vacuum field. Our main conclusions are as follows.

\begin{enumerate}
    \item The \citetalias{Olmschenk2025} surrogate-based MCMC framework extends to \psrseven, a substantially more challenging target than PSR~J0030$+$0451 because of its lower signal-to-background ratio. The neural-network surrogate provides a speedup of $\gtrsim 400$, making exploration of the 11-dimensional magnetic-field parameter space practical.

    \item The offset dipole-plus-quadrupole model provides statistically acceptable fits to the bolometric \nicer\ profile. The posterior-predictive light curves are consistent with the data, and the reduced-$\chi^2$ distributions for the full offset runs are centered close to unity.

    \item The inferred posterior is broad and multimodal, reflecting both the observational uncertainties of \psrseven\ and the degeneracies of the multipolar field parameterization. This breadth quantifies the range of magnetic topologies that remain compatible with the current bolometric data.

    \item The results are stable against the two calibrated temperature-weight prescriptions, $T_w=1.31$ and $T_w=1.41$, which encode the relative bolometric weighting associated with the effective-temperature difference between the two emitting regions.

    \item Within the restricted field family considered here, the data favor the offset model over a centered dipole plus centered axisymmetric quadrupole. The zero-offset run yields reduced-$\chi^2$ values systematically larger than those of the offset runs and generally above unity, although this should not be interpreted as proof that the physical moments must be displaced from the stellar center, because a more general centered multipolar expansion could provide additional flexibility.

    \item The inferred hotspot coverage is broad but structured. The pulse phases constrain the approximate azimuthal placement of the emitting regions, while the bolometric light curve leaves substantial freedom in latitude, area, and footpoint morphology. This broad extent remains notable despite fixing $M$, $R$, and the observer inclination to the highest-likelihood values of \citet{Dittmann2024}.

    \item The magnetic-moment direction maps suggest that the dipole and quadrupole components play different roles in shaping the surface footpoints. The dipole directions tend to lie near the inferred hotspot regions, whereas the quadrupole directions are more concentrated in the northern rotational hemisphere, with additional lower-latitude solutions along similar azimuths.

    \item The main limitations of the present analysis are the static-vacuum field approximation, the use of a bolometric-only light curve, the simplified atmosphere/beaming prescription, and the restricted rotated $m=0$ quadrupole. Extensions incorporating energy-dependent pulse profiles, more realistic force-free magnetic-field structures, and joint X-ray/$\gamma$-ray light-curve fitting are currently under development and will be presented in future studies.
\end{enumerate}


\begin{acknowledgments}
The material is based upon work supported by NASA under award numbers 80GSFC24M0006, 80GSFC21M0002, and 80NSSC21K1999, and under grants 21-ATP21-0116, 22-ADAP22-0142, 22-TCAN22-0027. 
F.T. and S.D. were also supported by the Laboratory Directed Research and Development program of Los Alamos National Laboratory project 20250750ECR. A.B.S. and D.O. were supported by the Laboratory Directed Research and Development program of Los Alamos National Laboratory project 20250637DI.
W.F.W. acknowledges funding by the Deutsche Forschungsgemeinschaft (DFG, German Research Foundation) under Germany’s Excellence Strategy – EXC 2121 “Quantum Universe” – 390833306.
Resources supporting this
work were provided by the NASA High-End Computing (HEC)
Program through the NASA Advanced Supercomputing (NAS)
Division at Ames Research Center. We acknowledge the use of resources of the National Energy Research Scientific Computing Center, a DOE Office of Science User Facility supported by the Office of Science of the U.S. Department of Energy under Contract No. DE-AC02-05CH11231 using NERSC award NP-ERCAP0037206.
This research also used resources provided by the Los Alamos National Laboratory Institutional Computing Program, which is supported by the U.S. Department of Energy National Nuclear Security Administration under Contract No. 89233218CNA000001. 
We also acknowledge the use of the NASA Astrophysics Data Service (ADS).
\end{acknowledgments}








\bibliographystyle{aasjournalv7}



\bibliography{main}

@ARTICLE{Raaijmakers2021,
  author  = {{Raaijmakers}, G. and {Greif}, S.~K. and {Hebeler}, K. and
             {Hinderer}, T. and {Nissanke}, S. and {Schwenk}, A. and
             {Riley}, T.~E. and {Watts}, A.~L. and {Lattimer}, J.~M. and
             {Ho}, W.~C.~G.},
  title   = "{Constraints on the Dense Matter Equation of State and Neutron
              Star Properties from NICER's Mass--Radius Estimate of
              PSR J0740+6620 and Multimessenger Observations}",
  journal = {ApJL},
  year    = {2021},
  volume  = {918},
  pages   = {L29},
  doi     = {10.3847/2041-8213/ac089a},
  archivePrefix = {arXiv},
  eprint  = {2105.06981},
  primaryClass = {astro-ph.HE}
}

@ARTICLE{Salmi2024,
       author = {{Salmi}, Tuomo and {Choudhury}, Devarshi and {Kini}, Yves and {Riley}, Thomas E. and {Vinciguerra}, Serena and {Watts}, Anna L. and {Wolff}, Michael T. and {Arzoumanian}, Zaven and {Bogdanov}, Slavko and {Chakrabarty}, Deepto and {Gendreau}, Keith and {Guillot}, Sebastien and {Ho}, Wynn C.~G. and {Huppenkothen}, Daniela and {Ludlam}, Renee M. and {Morsink}, Sharon M. and {Ray}, Paul S.},
        title = "{The Radius of the High-mass Pulsar PSR J0740+6620 with 3.6 yr of NICER Data}",
      journal = {\apj},
         year = 2024,
        month = oct,
       volume = {974},
       number = {2},
          eid = {294},
        pages = {294},
          doi = {10.3847/1538-4357/ad5f1f},
archivePrefix = {arXiv},
       eprint = {2406.14466},
 primaryClass = {astro-ph.HE},
       adsurl = {https://ui.adsabs.harvard.edu/abs/2024ApJ...974..294S}
}

@ARTICLE{Kundu2026,
  author  = {{Kundu}, A. and {Kalapotharakos}, C. and {Wadiasingh}, Z. and
             {Olmschenk}, G. and {Wallace}, W.~F. and {Harding}, A.~K. and
             {Venter}, C. and {Kazanas}, D.},
  title   = "{The swept-back multipolar magnetic field of neutron stars:
              Application to NICER MSP J0030+0451}",
  journal = {arXiv e-prints},
  year    = 2026,
  eprint  = {2604.19534},
  archivePrefix = {arXiv},
  primaryClass  = {astro-ph.HE}
}

@article{GoodmanWeare2010,
  author  = {Goodman, Jonathan and Weare, Jonathan},
  title   = {Ensemble samplers with affine invariance},
  journal = {Communications in Applied Mathematics and Computational Science},
  year    = {2010},
  volume  = {5},
  number  = {1},
  pages   = {65--80},
  doi     = {10.2140/camcos.2010.5.65}
}

@misc{Olmschenk2025haplo,
  author    = {Olmschenk, Greg and Kalapotharakos, Constantinos and
               Wadiasingh, Zorawar and Broadbent, Emily and
               Wallace, Wendy Fu},
  title     = {{haplo}},
  year      = {2025},
  publisher = {Zenodo},
  doi       = {10.5281/zenodo.14814679},
  url       = {https://doi.org/10.5281/zenodo.14814679},
  note      = {Version v1}
}

@inproceedings{He2016,
  author    = {He, Kaiming and Zhang, Xiangyu and Ren, Shaoqing and Sun, Jian},
  title     = {Deep Residual Learning for Image Recognition},
  booktitle = {Proceedings of the IEEE Conference on Computer Vision and
               Pattern Recognition (CVPR)},
  year      = {2016},
  pages     = {770--778},
  doi       = {10.1109/CVPR.2016.90}
}

@inproceedings{He2015,
  author    = {He, Kaiming and Zhang, Xiangyu and Ren, Shaoqing and Sun, Jian},
  title     = {Delving Deep into Rectifiers: Surpassing Human-Level Performance
               on {ImageNet} Classification},
  booktitle = {Proceedings of the IEEE International Conference on Computer
               Vision (ICCV)},
  year      = {2015},
  pages     = {1026--1034},
  doi       = {10.1109/ICCV.2015.123}
}

@inproceedings{LoshchilovHutter2019,
  author    = {Loshchilov, Ilya and Hutter, Frank},
  title     = {Decoupled Weight Decay Regularization},
  booktitle = {International Conference on Learning Representations (ICLR)},
  year      = {2019},
  url       = {https://openreview.net/forum?id=Bkg6RiCqY7}
}

@ARTICLE{Alpar1982,
  author  = {{Alpar}, M.~A. and {Cheng}, A.~F. and {Ruderman}, M.~A. and
             {Shaham}, J.},
  title   = "{A new class of radio pulsars}",
  journal = {Nature},
  year    = 1982,
  volume  = {300},
  pages   = {728--730},
  doi     = {10.1038/300728a0}
}

@ARTICLE{BhattacharyaVDH1991,
  author  = {{Bhattacharya}, D. and {van den Heuvel}, E.~P.~J.},
  title   = "{Formation and evolution of binary and millisecond
              radio pulsars}",
  journal = {Phys. Rep.},
  year    = 1991,
  volume  = {203},
  pages   = {1--124},
  doi     = {10.1016/0370-1573(91)90064-S}
}

@ARTICLE{HardingMuslimov2001,
  author  = {{Harding}, A.~K. and {Muslimov}, A.~G.},
  title   = "{Pulsar polar cap heating and surface thermal X-ray
              emission. I.}",
  journal = {ApJ},
  year    = 2001,
  volume  = {556},
  pages   = {987--1001},
  doi     = {10.1086/321589}
}

@ARTICLE{HardingMuslimov2002,
  author  = {{Harding}, A.~K. and {Muslimov}, A.~G.},
  title   = "{Pulsar polar cap heating and surface thermal X-ray
              emission. II.}",
  journal = {ApJ},
  year    = 2002,
  volume  = {568},
  pages   = {862--877},
  doi     = {10.1086/338985}
}

@ARTICLE{Contopoulos1999,
  author  = {{Contopoulos}, I. and {Kazanas}, D. and {Fendt}, C.},
  title   = "{The axisymmetric pulsar magnetosphere}",
  journal = {ApJ},
  year    = 1999,
  volume  = {511},
  pages   = {351--358},
  doi     = {10.1086/306652}
}

@ARTICLE{BaiSpitkovsky2010,
  author  = {{Bai}, X.-N. and {Spitkovsky}, A.},
  title   = "{Modeling of gamma-ray pulsar light curves using the
              force-free magnetic field}",
  journal = {ApJ},
  year    = 2010,
  volume  = {715},
  pages   = {1282--1301},
  doi     = {10.1088/0004-637X/715/2/1282}
}

@INPROCEEDINGS{Gendreau2016,
  author       = {{Gendreau}, K.~C. and {Arzoumanian}, Z. and
                  {Adkins}, P.~W. and others},
  title        = "{The Neutron star Interior Composition Explorer
                   (NICER): design and development}",
  booktitle    = {Proc. SPIE},
  year         = 2016,
  volume       = {9905},
  pages        = {99051H},
  doi          = {10.1117/12.2231304}
}

@ARTICLE{Bogdanov2019,
  author  = {{Bogdanov}, S. and {Guillot}, S. and {Ray}, P.~S. and others},
  title   = "{Constraining the neutron star mass-radius relation and
              dense matter equation of state with NICER. I. The
              millisecond pulsar PSR J0030+0451}",
  journal = {ApJL},
  year    = 2019,
  volume  = {887},
  pages   = {L25},
  doi     = {10.3847/2041-8213/ab53eb}
}

@ARTICLE{Miller2019,
  author  = {{Miller}, M.~C. and {Lamb}, F.~K. and {Dittmann}, A.~J. and
             others},
  title   = "{PSR J0030+0451 mass and radius from NICER data and
              implications for the properties of neutron star matter}",
  journal = {ApJL},
  year    = 2019,
  volume  = {887},
  pages   = {L24},
  doi     = {10.3847/2041-8213/ab50c5}
}

@ARTICLE{Riley2019,
  author  = {{Riley}, T.~E. and {Watts}, A.~L. and {Bogdanov}, S. and others},
  title   = "{A NICER view of PSR J0030+0451: Millisecond pulsar
              parameter estimation}",
  journal = {ApJL},
  year    = 2019,
  volume  = {887},
  pages   = {L21},
  doi     = {10.3847/2041-8213/ab481c}
}

@ARTICLE{Bilous2019,
  author  = {{Bilous}, A.~V. and {Watts}, A.~L. and {Harding}, A.~K. and
             others},
  title   = "{A NICER view of PSR J0030+0451: Evidence for a
              global-scale multipolar magnetic field}",
  journal = {ApJL},
  year    = 2019,
  volume  = {887},
  pages   = {L23},
  doi     = {10.3847/2041-8213/ab53e7}
}

@ARTICLE{Chen2020,
  author  = {{Chen}, A.~Y. and {Yuan}, Y. and {Vasilopoulos}, G.},
  title   = "{Multipolar magnetic fields in millisecond pulsars}",
  journal = {ApJL},
  year    = 2020,
  volume  = {893},
  pages   = {L38},
  doi     = {10.3847/2041-8213/ab85c5}
}

@ARTICLE{Cromartie2020,
  author  = {{Cromartie}, H.~T. and {Fonseca}, E. and {Ransom}, S.~M. and
             others},
  title   = "{Relativistic Shapiro delay measurements of an
              extremely massive millisecond pulsar}",
  journal = {Nat. Astron.},
  year    = 2020,
  volume  = {4},
  pages   = {72--76},
  doi     = {10.1038/s41550-019-0880-2}
}

@ARTICLE{Fonseca2021,
  author  = {{Fonseca}, E. and {Cromartie}, H.~T. and {Pennucci}, T.~T. and
             others},
  title   = "{Refined mass and geometric measurements of the
              high-mass PSR J0740+6620}",
  journal = {ApJL},
  year    = 2021,
  volume  = {915},
  pages   = {L12},
  doi     = {10.3847/2041-8213/ac03b8}
}

@ARTICLE{Miller2021,
  author  = {{Miller}, M.~C. and {Lamb}, F.~K. and {Dittmann}, A.~J. and
             others},
  title   = "{The radius of PSR J0740+6620 from NICER and XMM-Newton data}",
  journal = {ApJL},
  year    = 2021,
  volume  = {918},
  pages   = {L28},
  doi     = {10.3847/2041-8213/ac089b}
}

@ARTICLE{Riley2021,
  author  = {{Riley}, T.~E. and {Watts}, A.~L. and {Ray}, P.~S. and others},
  title   = "{A NICER view of the massive pulsar PSR J0740+6620
              informed by radio timing and XMM-Newton spectroscopy}",
  journal = {ApJL},
  year    = 2021,
  volume  = {918},
  pages   = {L27},
  doi     = {10.3847/2041-8213/ac0a81}
}

@ARTICLE{Dittmann2024,
  author  = {{Dittmann}, A.~J. and {Miller}, M.~C. and {Lamb}, F.~K. and
             others},
  title   = "{A more precise measurement of the radius of
              PSR J0740+6620 from updated NICER data}",
  journal = {ApJ},
  year    = 2024,
  volume  = {974},
  pages   = {295},
  doi     = {10.3847/1538-4357/ad5f1e}
}

@ARTICLE{Kalapotharakos2014,
  author  = {{Kalapotharakos}, C. and {Harding}, A.~K. and {Kazanas}, D.},
  title   = "{Gamma-ray emission in dissipative pulsar magnetospheres:
              from theory to Fermi observations}",
  journal = {ApJ},
  year    = 2014,
  volume  = {793},
  pages   = {97},
  doi     = {10.1088/0004-637X/793/2/97}
}

@ARTICLE{Kalapotharakos2021,
  author  = {{Kalapotharakos}, C. and {Wadiasingh}, Z. and {Harding}, A.~K. and
             {Kazanas}, D.},
  title   = "{The multipolar magnetic field of the millisecond pulsar
              PSR J0030+0451}",
  journal = {ApJ},
  year    = 2021,
  volume  = {907},
  pages   = {63},
  doi     = {10.3847/1538-4357/abcec0}
}

@ARTICLE{Olmschenk2025,
  author  = {{Olmschenk}, G. and {Broadbent}, E. and {Kalapotharakos}, C. and
             {Wallace}, W.~F. and {Lechien}, T. and {Wadiasingh}, Z. and
             {Kazanas}, D. and {Harding}, A.},
  title   = "{Pioneering high-speed pulsar parameter estimation using
              convolutional neural networks}",
  journal = {ApJ},
  year    = 2025,
  volume  = {991},
  pages   = {169},
  doi     = {10.3847/1538-4357/ae03c0}
}

@ARTICLE{Petri2023,
  author  = {{P{\'e}tri}, J. and {Guillot}, S. and {Guillemot}, L. and
             others},
  title   = "{Joint X-ray, radio, and gamma-ray fitting of PSR J0030+0451}",
  journal = {A\&A},
  year    = 2023,
  volume  = {680},
  pages   = {A93},
  doi     = {10.1051/0004-6361/202346913}
}

@article{Lechien2026,
  author  = {{Lechien}, T. and {Olmschenk}, G. and 
             {Kalapotharakos}, C. and others},
  title   = {Force-free pulsar magnetosphere inference via 
             neural-network surrogates},
  year    = {2026},
  journal = {in preparation}
}

\end{document}